\documentclass[11pt,a4paper]{article}
\pdfoutput=1
\usepackage{jcappub}

\usepackage{amsmath, amsthm, amssymb, amsfonts}
\usepackage{slashed}
\usepackage[ngerman,english]{babel}
\usepackage{epsfig, graphicx}
\usepackage{color}
\usepackage{lscape}
\usepackage{chngcntr}
\usepackage{threeparttable}

\newcommand{\U}{\ensuremath{U(1)_\text{h}}}

\newcommand{\dd}{\text{d}}
\newcommand{\gt}{\ensuremath{\Gamma_{\gamma^{*}}}}

\newcommand{\Pt}{\ensuremath{\Pi}}

\newcommand{\apss}{Ap\&SS}

\newcommand{\prd}{Phys. Rev. D}

\newcommand{\ssr}{Space Sci. Rev.}
\newcommand{\araa}{ARA\&A}
\newcommand{\apj}{ApJ}
\newcommand{\apjl}{ApJL}
\newcommand{\mnras}{MNRAS}
\newcommand{\jcap}{JCAP}
\newcommand{\aap}{A\& A}
\newcommand{\nat}{Nature}

\begin{document}

\subheader{\hfill MPP-2015-199}

\title{Minicharged Particles from the Sun: A Cutting-Edge Bound}

\author[a]{N.~Vinyoles,}
\author[b]{H.~Vogel}

\affiliation[a]{Institute of Space Sciences (CSIC-IEEC),\\
Campus UAB, Carrer de Can Magrans, S/N, 08193 Cerdanyola del Vall\`es, Spain}
\affiliation[b]{Max Planck Institute for Physics (Werner Heisenberg Institute), \\ 
F\"ohringer Ring 6, D-80805 M\"unchen, Germany}

\emailAdd{vinyoles@ice.csic.es}
\emailAdd{hvogel@mpp.mpg.de}

\abstract{We study the impact on the Sun of an exotic energy-loss channel caused by plasmon decay into fermionic minicharged particles with charge $\epsilon e$ and mass $m_f$. We compare solar models with this extra emission to helioseismological and neutrino data, obtaining a bound $\epsilon <2.2\times 10^{-14}\ ({\rm 95\%\, CL})$ for $m_f\lesssim 25 \, \text{eV}$. Our result is comparable to previous limits from the cooling of globular cluster stars, while at the same time it is better understood and takes theoretical and observational errors into account.
}

\maketitle

\section{Introduction}

Our Sun is a powerful laboratory. Its large density, temperature and shear size allow us to study physics in environments that are hard to reproduce in earth-based experiments. Its proximity, on the other hand, allows us to measure and understand the Sun in great detail. Major breakthroughs have been achieved in the past by observing, e.g., neutrinos from proton--proton fusion~\cite{Smirnov:2015lxy} or neutrino flavor oscillations~\cite{Ahmad:2002jz}.

The Sun is the most studied star and information from observations and theoretical models are available to look into its inside. Helioseismology~\cite{helioseismology}, the study of the global oscillations of the Sun, allows us to extract information on the Sun's structure, e.g., the sound speed profile, the convective radius and the surface helium abundance. Moreover, the measurement of the solar neutrino fluxes enables us to extract information on the solar interior through an independent channel. These observations combined with the accurate Standard Solar Models (SSMs)~\cite{bahcall2001,serenelli11} give us the key to understand the Sun's structure and evolution. Although some inconsistency remains~\cite{basu2004b, serenelli09}, the observations and solar models broadly agree.

This agreement allows us to constrain any contribution that might arise from non-standard physics. Weakly-interacting, low-mass particles like the axion~\cite{Peccei:1977hh} and axion-like particles (ALPs), hidden photons~\cite{Galison:1983pa, Holdom:1986eq, Okun:1982xi}, or minicharged particles (MCPs), e.g.~\cite{Holdom:1986eq, Foot:1989fh}, can be produced copiously inside the Sun. If the couplings of the exotic particles are so large that they are trapped, they contribute to the Sun's energy transport, potentially changing its composition and evolution drastically. For smaller couplings, they carry energy out of the Sun which changes the solar evolutionary structure. By comparing this structure with observations, one can derive stringent bounds on extra energy sinks. 
This energy-loss argument has been applied a number of times to the cooling of WDs~\cite{isern1992, davidson2000, isern2008, isern2010, Bertolami:2014rta} and supernovae (SNe)~\cite{Mohapatra:1990vq, Ayala:1998qz}, the start of helium burning in RGs~\cite{davidson2000, viaux2014, Redondo:2013lna,ayala2014,An:2014twa}, the lifetime of horizontal branch (HB) stars~\cite{davidson2000,viaux2014, raffelt1999, ayala2014, An:2013yfc,Redondo:2013lna,An:2014twa} and also to the evolution and composition of the Sun~\cite{raffelt1999, Vinyoles:2015aba, Redondo:2008aa,An:2014twa,Redondo:2013lna,Redondo:2013wwa}. Our Sun used to be much less constraining than the more extreme objects since it is less dense and hot. However, we will show that this detriment is compensated by the quality and amount of data available.

In this work, we focus on the energy-loss argument for minicharged particles produced in the Sun. Although the charge of all observed particles seems to be quantized and a multiple of the down-quark's charge, there is no a priori theoretical reason for charge quantization in the particle Standard Model (SM). This mystery has led to a continuous effort of experimental searches~\cite{Davidson:1991si,Gninenko:2006fi,Badertscher:2006fm, Prinz:1998ua, Jaeckel:2012yz, Haas:2014dda,Izaguirre:2015eya}. 
On the theoretical side, the missing quantization allows theorists to postulate the existence of a particle that couples to the SM photon with an arbitrary charge, the MCP. Such a particle can either be studied as an ad hoc extension of the SM, or it might be predicted by more intricate models, for example theories with an additional unbroken \U\ gauge symmetry~\cite{Holdom:1986eq, Holdom:1985ag} or even larger hidden sectors (e.g.~\cite{Foot:2014uba,Farina:2015uea}). 
The charge of the MCP is usually parametrized as $\epsilon e$, where $\epsilon$ is the minicharge parameter ('minicharge') and $e$ is the charge of the electron. The size of the 
minicharge depends on the model assumptions and can be very small (e.g. ~\cite{Burgess:2008ri,Blumenhagen:2005ga,Abel:2008ai,Bullimore:2010aj,Heckman:2010fh,Goodsell:2011wn}). We will in the following consider $\epsilon$ as a free parameter.

Minicharged particles couple to the electromagnetic plasma inside the Sun, which contains a thermal bath of collective excitation modes ('plasmons'). These plasmons can deexcite, thereby emitting a pair of MCPs. 
The rate of this decay depends on the minicharge but also on the plasma frequency $\omega_p$  of the Sun and the MCP mass $m_f$ since the decay into MCPs becomes kinematically disfavored for large MCP masses, $2m_f>\omega_p$. Once produced, the small minicharges we consider here allow the MCPs  to escape the Sun unimpeded, giving rise to an additional energy-loss channel.

 \begin{figure}[!t] \centering
\includegraphics[width=0.49\textwidth]{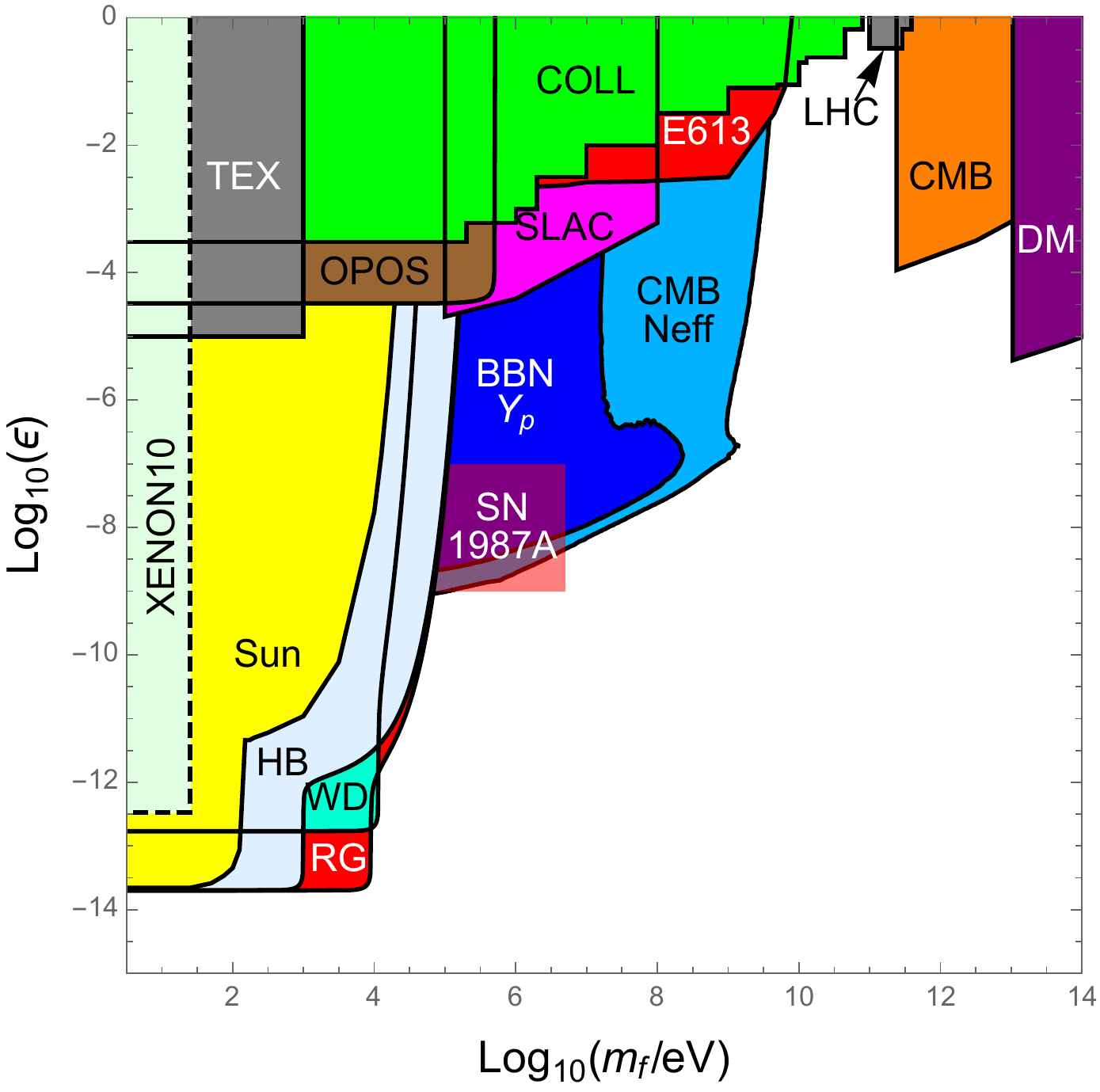}
\includegraphics[width=0.49\textwidth]{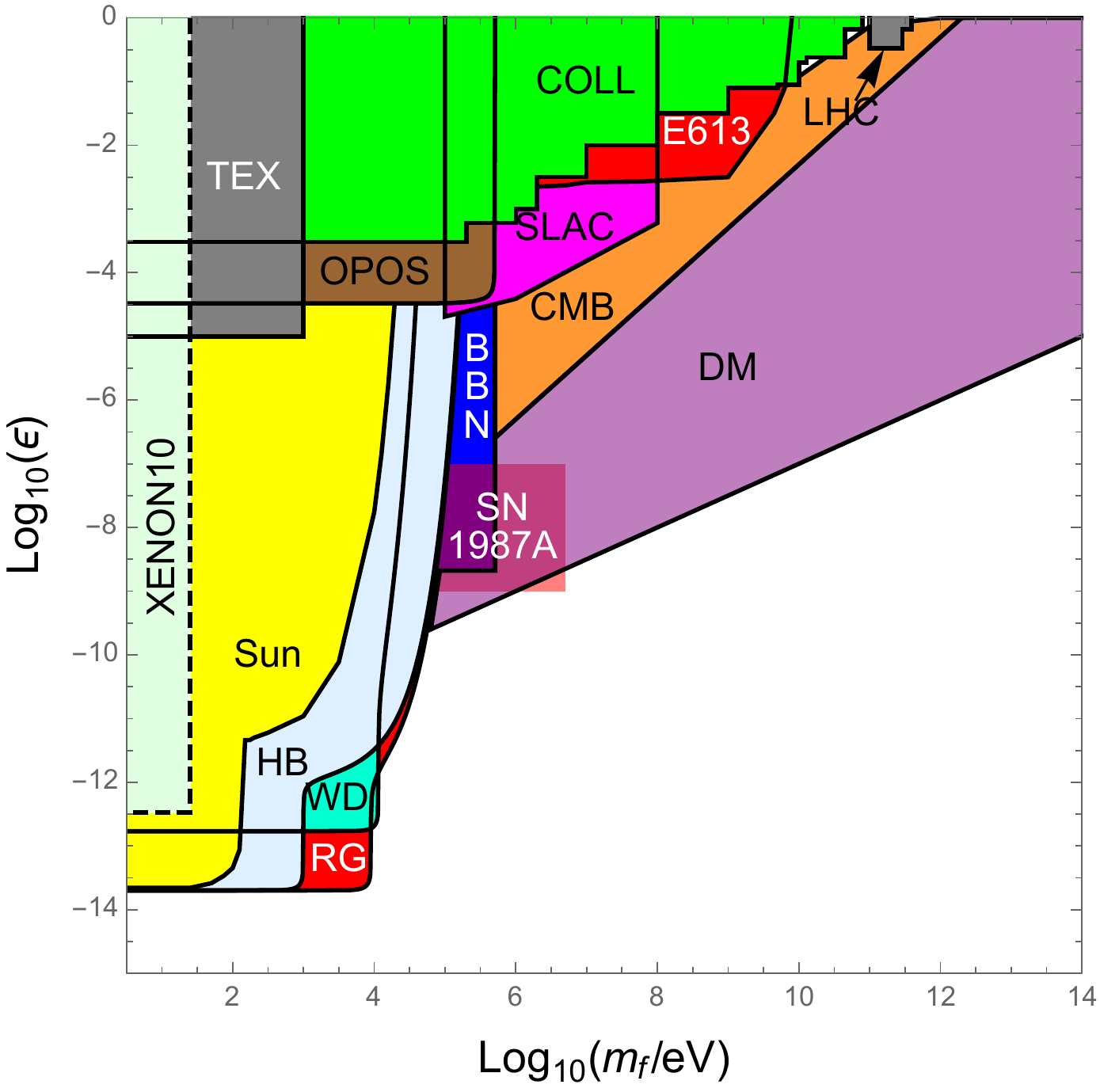}
 \caption[Exclusion plot for MCPs]{{\small{
Summary of constraints on fermionic MCPs in the mass/minicharge plane. The result of this work is depicted in yellow. \emph{Left panel:} Models with a massless hidden photon for a \U\ gauge coupling $g'=0.1$ (see sec.~\ref{sec:MCP}).  We updated the CMB and BBN bounds from ref.~\cite{Vogel:2013raa} using 2015 Planck data~\cite{Ade:2015xua}, and the collider bounds (COLL) from refs.~\cite{Davidson:1991si,Davidson:2000hf}. The remaining bounds are from DM~\cite{Davidson:1991si}, CMB~\cite{Dubovsky:2003yn}, LHC~\cite{Jaeckel:2012yz}, 
SLAC~\cite{Prinz:1998ua}, OPOS~\cite{Badertscher:2006fm}, TEX~\cite{Gninenko:2006fi}, E613~\cite{Soper:2014ska}, and HB, WD, RG~\cite{davidson2000, Vogel:2013raa}. The Xenon10 bound is plotted with dashed lines since it is an estimate taken from a higgsed model of minicharged particles~\cite{An:2013yua}.} \emph{Right panel:} Bounds on models without a hidden photon. The dark radiation bounds from ref.~\cite{Vogel:2013raa} 
disappear but the limits from overproduction~\cite{Davidson:1991si} and distortions of the CMB~\cite{Dubovsky:2003yn} 
become far more constraining.}} \label{fig:MCPresult} \end{figure}
To limit this emission from the Sun, we use the statistical method described in ref.~\cite{Vinyoles:2015aba} that combines most of the available information on the Sun: observational (neutrino fluxes and helioseismology) and theoretical (SSMs) data taking into account the corresponding errors and correlations. This leads to a conservative bound (95\% CL) of 

\begin{equation}
 \epsilon < 2.2\times \smash 10^{-14} \quad{\rm for}\ m_f < 25\, \text{eV}\,,
\end{equation}
 on the minicharge for low-mass MCPs, which is comparable to the limits $\epsilon < 2\times 10^{-14}$ obtained by red giants and horizontal branch stars~\cite{davidson2000}. Equivalently, we find that an emission of up to $1.5\%$ of the Sun's total luminosity can still be viable for these MCPs.
 
The exact mass dependence of our bound can be seen in fig.~\ref{fig:MCPresult} together with a summary of limits that have been obtained previously from various experiments and observations\footnote{Figure~\ref{fig:MCPresult} also includes updated limits from dark radiation~\cite{Vogel:2013raa} and colliders~\cite{Davidson:1991si, Davidson:2000hf}.}. The plots show the two most popular scenarios that involve MCPs. The bottom-up approach minimally adds an additional particle with arbitrary charge, the MCP, to the Standard Model. This is plotted in the right panel of fig.~\ref{fig:MCPresult}. 
Theoretically more motivated scenarios generate a minicharge through the kinetic mixing of the SM photon with a new massless photon. Some bounds on MCPs rely on this new 'hidden' photon, so that the summary plot changes drastically for large MCP masses. The bounds for this model are shown in the left panel of fig.~\ref{fig:MCPresult}. Our limit is identical for both models and is displayed in yellow. As apparent from the 
figure, the bound of this work is among the most constraining for low-mass MCPs.  

Our result is especially interesting because the bound is more robust than previous limits and contains statistical information that is absent for bounds obtained from RGs and HB stars. Additionally, our Sun is not only better observed than any other astrophysical object, future improvement of our understanding is to be expected as well. Additional insight can be achieved by the new generation of neutrino experiments, the possibility of observing g-modes~\cite{garcia2007}, which will give direct information on the center of the Sun, 
or new opacity experiments (e.g.~\cite{bailey2015}) among other improvements. 
This will very likely make the Sun the most constraining source for low-mass MCPs in the foreseeable future.

The plan of the paper is as follows. We present the observational and theoretical status of the Sun in sec.~\ref{sec:sunbasics}. Minicharged particles, their analytical emission rates and some comments on their propagation can be found in sec.~\ref{sec:MCP}. This is proceeded in sec.~\ref{sec:statistical} by the statistical method applied in this paper. We then introduce the MCP into the SSMs in sec.~\ref{sec:MCPSun} and comment on changes of the observables. The values of our resulting bounds on $\epsilon$ are presented in sec.~\ref{sec:results}. The discussion and summary can be found in sec.~\ref{sec:conclusion}.

\section{The Sun}\label{sec:sunbasics}

The key advantage in using the Sun lies in the amount of available information. Our goal is to combine the helioseismology data and the neutrino solar observations (see sec.~\ref{sec:Obs}) with  solar models (see sec.~\ref{sec:SSM}), in order to place upper limits on the properties of the minicharged particles. In this section, we describe the available information on the Sun that we use for this purpose.

\subsection{Observables}
\label{sec:Obs}
 The most important information on the Sun comes from helioseismology and from the detected neutrino fluxes. Helioseismology allows us to determine the sound speed profile. This can be done by inverting the observed  frequencies of solar oscillations, providing a direct test of the structure of the solar interior. From helioseismology, we can also derive the surface helium abundance and the radius of the convective zone~\cite{helioseismology}. In particular, we use as observables 30 points of the sound speed profile with radii $r<0.8 R_\odot$ from ref.~\cite{basu2007}, the surface
helium $Y_S$ and the convective radius $R_{\text{CZ}}$. From the neutrino observations, we consider the neutrino fluxes from $^8\rm{B}$ and $^7 \rm{Be}$. We do not use the $pp$ and $pep$ neutrino fluxes because the experimental errors are still too large to be statistically relevant. The observational values and the corresponding errors of the surface helium, radius of the convective zone and the neutrino fluxes are the same values as used in ref.~\cite{Vinyoles:2015aba} and are summarized in tab.~\ref{tab:obs_value} together with the values of the AGSS09 solar model and the corresponding theoretical errors.

\begin{table}[t]
\begin{center}
\begin{tabular}{ | c c c c |}\hline
 & AGSS09 & Observables & Ref. \\
\hline
$\rm{Y_s}$ & $0.232 (1 \pm 0.013)$ &$0.2485 \pm 0.0035$& \cite{basu1997},\cite{basu2004b}\\
$\rm{R_{CZ}/R_{\odot}}$ & $0.7238 (1 \pm 0.0033) $& $0.713 \pm 0.001$& \cite{basu1997},\cite{basu2004b}\\
$\rm{\Phi(^7Be)}$ & $4.56 (1  \pm 0.06) $ & $4.82(1^{+0.05}_{-0.04})$ & \cite{bellini2011}, \cite{gonzalez2010}\\
$\rm{\Phi(^8B})$ & $4.60 (1 \pm 0.11)$ &$5.00(1 \pm 0.03)$& \cite{bellini2011},\cite{gonzalez2010}\\
\hline
\end{tabular}
\caption{\small Theoretical and observational values for the surface helium $Y_\text{s}$, the convective radius $R_{\text{CZ}}$, the neutrino flux from $^7 \rm{Be}$ and from $^8 \rm{B}$. The second column shows the SSM values for the AGSS09 composition with the corresponding model errors~\cite{villante2014}. The third column summarizes the observational and experimental values and errors. Neutrino fluxes are in $10^9\, \rm{cm^{-2} s^{-1}}$ for $\rm{\Phi(^7Be)}$ and $10^6 \,\rm{cm^{-2} s^{-1}}$ for $\rm{\Phi(^8B)}$.
\label{tab:obs_value}}
\end{center}
\end{table}

\subsection{Standard Solar Models}
\label{sec:SSM}

Standard Solar Models are theoretical descriptions of the Sun that are calibrated to match the Sun's present status. They are calculated by adjusting three initial quantities (initial helium, mixing length and initial metallicity) to satisfy the present solar luminosity $L_\odot = 3.8418 \times 10^{33}\, \rm{erg\ s^{-1}}$, the solar radius $R_\odot = 6.9598 \times10^{10}\, \rm{cm}$ and the metal-hydrogen ratio $(Z/X)_\odot$. This last ratio will determine the composition and distribution of metals in the interior of the Sun. Nowadays, the solar value of this ratio is still under debate. 
The main discussion is among two different compositions: The old compilation of abundances, GS98~\cite{gs98ab}, gives a value of $Z/X_\odot = 0.0229$ and a resulting thermal stratification close to observation. On the other hand, the new composition AGSS09~\cite{agss09ab} has been extracted using a more accurate analysis with 3D hydrodynamical models of the solar atmosphere and an improved solar spectroscopic analysis giving a metal-hydrogen ratio $(Z/X)_\odot=0.0178$. However, the resulting thermal stratification is worse compared to observations than for the older composition. This is the so-called \textit{solar abundance problem}.
One reason for this discrepancy could be that the radiative opacities, that together with the composition define the thermal stratification of the Sun, should be revised~\cite{villante2014, cd2009}. A recent calculation of the iron opacity~\cite{bailey2015} gives a larger value than the one predicted for solar temperatures, which goes in the direction of relieving the solar abundance problem. For more details about the compositions and the solar abundance problem, we refer the reader to refs.~\cite{basu2004b, serenelli09}.

In this work, we use GARSTEC (GARching STEllar Code) \cite{GARSTEC} to calculate the SSMs. Details on the parameters used can be found in section 2.1 of ref.~\cite{Vinyoles:2015aba}. In order to accommodate MCPs, the code has to be modified to account for the extra energy carried away.  This energy-loss from MCPs will be explained in the next section.

\section{Minicharged particles}\label{sec:MCP}

The Standard Model of particle physics can be extended by consistently introducing new fields.
 In this work, we derive a bound for the minimal extension of just adding a minicharged field to the SM Lagrangian. However, our bounds are also applicable to more extended models like MCPs accompanied by a massless hidden photon and more intricate hidden sectors with very massive degrees of freedom.

\subsection{Models without a hidden photon}

The Standard Model gauge groups do not prevent us from adding charged, massive particles to our theory. Indeed, a $SU(2)_{\rm{L}}$ singlet, the MCP, can always be added to the SM Lagrangian $\mathcal{L}_{\text{SM}}$,
\begin{equation}
\mathcal{L}=\mathcal{L_{\text{SM}}} +\bar f (i\slashed \partial -m_f )f +\epsilon e A^{\mu}\bar f \gamma_{\mu} f\,,
\end{equation}
where $f$ denotes the minicharged fermion with a mass $m_f$ (the MCP), $\epsilon$ is the minicharge parameter ('minicharge') and $A^\mu$ is the SM photon's vector potential. The minicharge $\epsilon$ is given by the charge quantum number of the MCP, which can be arbitrary small. This charge leads to a coupling between the MCP and photons, which also allows plasmons inside the Sun to decay into MCPs.

\subsection{Models with a hidden photon}

Minicharged particles can be naturally obtained when a local, unbroken gauge group $\U$ is added to the SM groups~\cite{Holdom:1986eq, Holdom:1985ag}. Here, the subscript $h$ denotes 'hidden' meaning that SM particles will not be charged under the \U . The Lagrangian for this model reads
\begin{equation}\label{eq:kinetic}
\mathcal{L} = \mathcal{L_{\text{SM}}}+ \bar f (i\slashed{D}-m_f) f -\frac{1}{4}F'_{\mu \nu}F'^{\mu \nu} -\frac{\kappa}{2}F_{\mu \nu} F'^{\mu \nu}\,,
\end{equation}
where the \U\ manifests itself in a field strength tensor $F'_{\mu\nu}$ with a vector potential $A'_{\mu}$, which describes the hidden photon. Again, 'hidden' means that no SM particle couples to the hidden photon. In contrast, $F_{\mu \nu}$ is the field strength tensor of the SM photon. Both photons have identical quantum numbers such that they can mix kinetically with an approximate mixing angle $\kappa$ [last term of eq.~\eqref{eq:kinetic}]. Such a mixing should not be omitted in generic models, and can also arise from loops of heavy particles charged under both, hypercharge and the hidden \U~\cite{Holdom:1985ag}. The covariant derivative is
\begin{equation}
D_\mu=\partial_\mu -i g' A'_\mu\,,
\end{equation}
where $g'$ is the \U\ gauge coupling of the fermion $f$. We also assume that $f$ does not interact through other channels than the hidden photon.

The fermion $f$ manifestly becomes a MCP when the equations of motion of the photons are diagonalized via a redefinition $A'_\mu\to A'_\mu-\kappa A_\mu$ and a rescaling of $A_\mu$. This reparametrization induces a coupling of the fermion $f$ to the SM photon with a strength $g'\kappa = \epsilon e$, where $\epsilon$ absorbs the degeneracy of $g'$ and $\kappa$. We, again, obtain a massive fermion with a minicharge $\epsilon$, while the hidden photon decouples from any SM particle~\cite{Holdom:1985ag}.

\subsection{Production and propagation of minicharged particles} 

Light particles that are being produced inside the Sun might alter its evolution drastically. If these particles escape from the Sun, they carry away energy that would change the temperature profile and neutrino fluxes among others (see sec.~\ref{sec:MCPSun}). This might potentially conflict with observations. Should the MCPs not be able to escape the Sun freely due to scattering or magnetic fields, they will contribute significantly to the energy transport inside the Sun, changing its  
composition and temperature distribution even more radically. In the following, we will concentrate on MCPs that escape from the Sun unimpeded. Their production will be discussed now, followed by 
an estimate for the validity of the assumption that scattering of MCPs during their propagation can be neglected.

\subsubsection{Production}

MCPs are dominantly produced through plasmon decay $\gamma^*\to f\bar f$ while other processes like $e^+e^-$-annihilation ($e^+e^- \to f\bar f$) or vector boson fusion $\gamma \gamma' \to f \bar f$ are suppressed due to the small number of positrons and hidden photons. On the other hand, SM photon fusion $\gamma\gamma \to f \bar f$ is of higher order in  the small parameter $\epsilon$. Hence, in the remainder we will only consider plasmon decay.

Plasmons are excitations of the dense electron-proton plasma. Alternatively, they can be viewed as photons that propagate inside the plasma and obtain a non-trivial dispersion relation with an effective mass. There are two types of such excitations: transversal and longitudinal plasmons. Since the number density of longitudinal photons in the Sun is much smaller than the number of transversal photons due to phase space limitations ($\omega_p^3\ll T^3$), we will only consider transversal excitations in the remainder of this work (see~\cite{Raffelt:1996wa}).

The energy emission rate per volume $\mathcal{Q}=dE/(dV dt)$ for plasmons decaying into MCPs reads~\cite{Raffelt:1996wa}
\begin{align}\label{eq:rate}
 \mathcal{Q} = \frac{2}{2\pi^2}\int_0^\infty \dd k k^2\frac{ \omega \Gamma_{\gamma^{*}}}{e^{\omega/T}-1}\,,
\end{align}
where $T$ is the temperature of the Sun and the plasmon's frequency $\omega$ is related to its momentum $k$ through the  dispersion relation 
 $\omega^2 -k^2 = \omega_p^2$, which holds for a transverse plasmon in a non-relativistic, non-degenerate plasma~\cite{Raffelt:1996wa}. In this limit, the plasma frequency is given by
\begin{align}
 \omega_p^2 = \frac{4\pi\alpha n_e}{m_e}\,,
\end{align}
with the fine-structure constant $\alpha$, electron density $n_e$ and electron mass $m_e$.
The decay rate $\Gamma_{\gamma^{*}}$ reads
\begin{align}\label{eq:OnShell}
 \Gamma_{\gamma^{*}}=\epsilon^2 \frac{\alpha}{3} \frac{Z}{\omega}\left(\omega_p^2+2m_f^2\right)\sqrt{1-\frac{4m_f^2}{\omega_p^2}}\,,
\end{align}
  where the renormalization factor $Z$ is of order unity. Finally, a factor 2 in eq.~\eqref{eq:rate} arises because of the two different polarization states of transversal plasmons.

Equation~\eqref{eq:rate} is valid as long as $\omega_p \geq 2m_f$ so that plasmons that fulfill the dispersion relation ('on-shell' plasmons) can decay into MCPs. However, even when $\omega_p < 2m_f$, plasmonic excitations that are not on-shell ('off-shell' plasmons) produce MCPs. These off-shell plasmons are thermally distributed~\cite{Weldon:1983jn} and their emission rate can be obtained straight-forwardly~\cite{Vogel:2013raa},
\begin{align}\label{eq:rate2}
 \mathcal{Q}=&2\int_0^\infty \frac{k^2\dd k}{2\pi^2} \int_{\sqrt{4 m_f^2+k^2}}^\infty\frac{\omega \dd \omega}{\pi}\frac{2 {\rm Im}\,\Pt}{(K^2-{\rm Re}\,\Pt)^2+({\rm Im}\,\Pt)^2}\frac{\omega \gt(K^2) }{e^{\omega/T}-1}\,,
\end{align}
where $K^2=\omega^2 -k^2$, the decay rate is given by eq.~\eqref{eq:OnShell} with $\omega_p^2$ replaced by $K^2$, and the self energy is $\Pt=\omega_p^2 +i\omega \Gamma_{\text{Th}}$. The decay rate  for plasmons with $\omega_p < 2m_f$, $\Gamma_{\text{Th}}$, is controlled by Thomson scattering $\Gamma_{\text{Th}}=n_e\sigma_{\text{Th}}=n_e(8\pi\alpha^2)/(3m_e^2)$, as can be seen by analyzing the two-loop self energies of the photons~\cite{Weldon:1983jn}.

Using eqs.~\eqref{eq:rate} and~\eqref{eq:rate2}, we are able to compute the production of MCPs with non-zero masses in all areas of the Sun. On-shell decay eq.~\eqref{eq:rate} is valid in high density regions while eq.~\eqref{eq:rate2} can be used in low-density areas of the Sun.

\subsubsection{Propagation}

After being produced, MCPs start to propagate inside the dense solar medium. If the coupling strength of the MCPs is very weak, they will free-stream out of the Sun. For larger $\epsilon$ they will lose successively more energy during their propagation. In our computations, we assume that the MCPs do not lose energy while leaving the Sun, thereby carrying away all their initial energy.

We can estimate the validity of this assumption by comparing the MCPs' mean free path $\lambda$ to the radius of the Sun $R_\odot$. If $\lambda \lesssim R_\odot$, the probability for a MCP to scatter while leaving the Sun is high. On the other hand, such a scattering becomes more and more unlikely as $\lambda$ becomes larger than $R_\odot$.

 \begin{figure}[t] \centering
  \includegraphics[width=0.55\textwidth]{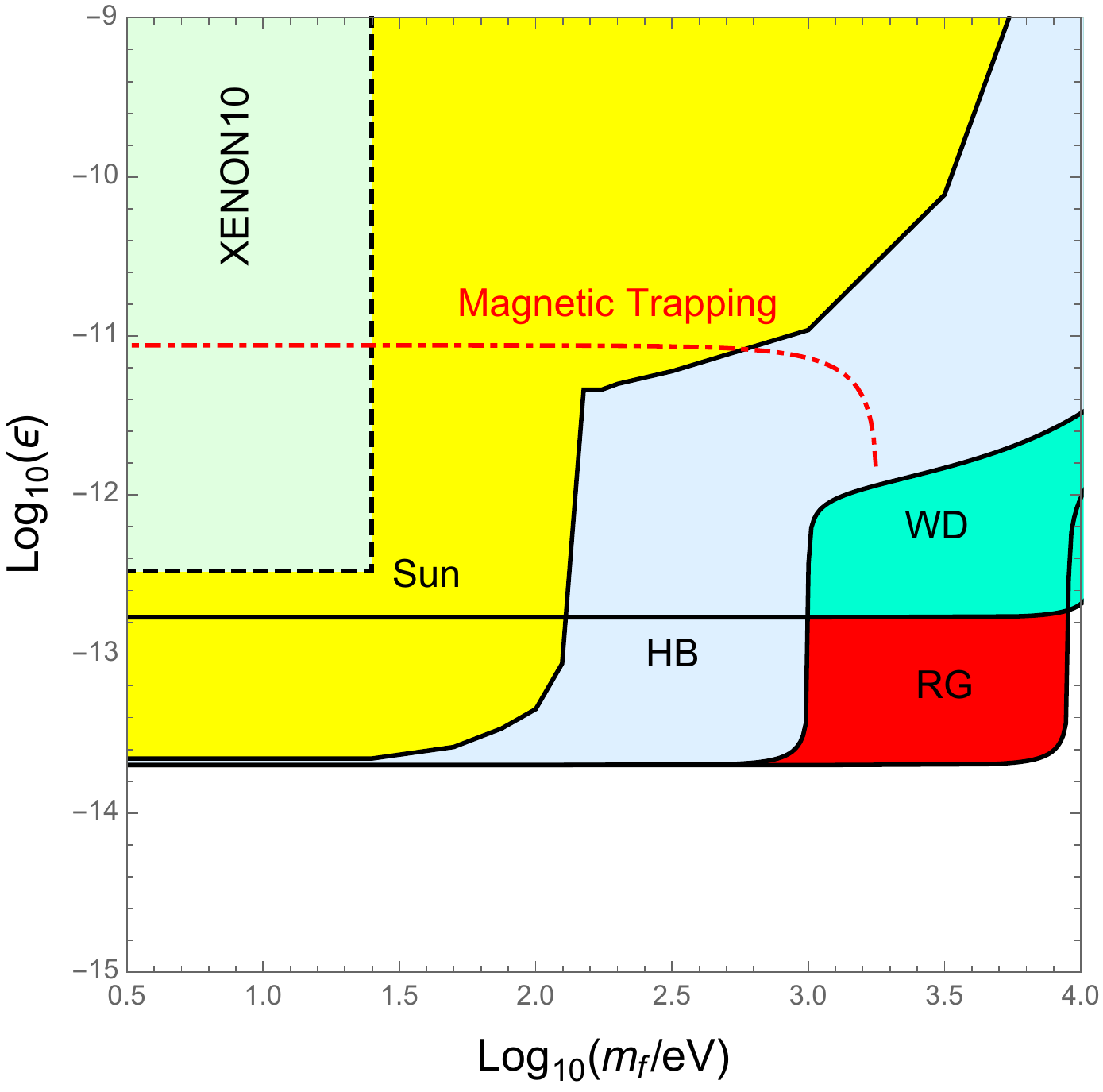}
 \caption[Exclusion plot for MCPs]{{\small{
Trapping of MCPs inside the Sun due to a magnetic field. For MCPs with large masses trapping becomes more effective and our assumptions of free streaming break down.}}} \label{fig:Trap} \end{figure}

The mean free path is controlled by Coulomb scattering on ambient electrons and protons. However, Coulomb scattering is strongly peaked in the forward direction in which MCPs are not deflected. Weighing the cross section with the deflected angle yields the more relevant transport cross section~\cite{davidson2000}
\begin{equation}
 \sigma_{\text{eff}}=\frac{2\pi\epsilon^2\alpha^2}{E^2}\left[\frac{2+z}{2}\text{ln}\left(\frac{2+z}{z}\right)-1\right]\,,
\end{equation}
where $z=k^2_D/(2E^2)$, which takes into account the screening effects via the Debye screening scale $k^2_D=4\pi\alpha n_{\rm{p, e}}/T$. Here, $n_p$ is the number density of protons, which approximately equals the number density of electrons, and $E\approx (3/2)T$ is the energy of the emitted MCPs, where the approximation holds for MCPs produced by typical thermal plasmons. In terms of this cross section, the mean free path reads
\begin{equation}
 \lambda = \frac{1}{2\sigma_{\text{eff}}n_e}\,,
\end{equation}
where we took into account electrons and protons by setting $ n_\text{e}\approx n_\text{p}$. Comparing this to the radius of the Sun $R_\odot\approx7\times 10^{10} \ \text{cm}$, we obtain that scattering of MCPs only becomes significant for
\begin{equation}
 \epsilon> 4.6\times 10^{-7}\,.
\end{equation}
Minicharged particles can also be trapped by magnetic fields if they force the MCPs on a Larmor radius that is of the order of the radius of the Sun. Even MCPs that leave the Sun might reenter while following a magnetic field line. 

The Larmor radius $r_\text{L}$ can be found to be
\begin{equation}
 r_\text{L}=\frac{\gamma\beta m_f}{\epsilon e B}= \frac{\sqrt{E^2-m_f^2}}{\epsilon eB}\,,
\end{equation}
where we take $B\approx 1\ \text{G}$ as the polar magnetic field of the Sun~\cite{lrsp-2010-1, Moss:2012yq}, and  $\gamma=E/m_f$ and $\beta = p/E$ are the usual parameters of a Lorentz boost with energy $E$ and momentum $p$. A comparison of this radius with 10 times the Sun's radius leads to a trapping for low-mass MCPs of the order
\begin{equation}
 \epsilon\gtrsim 8.8 \times 10^{-12}\,.
\end{equation}
The exact behavior with mass compared to the bound from fig.~\ref{fig:MCPresult} is shown in fig.~\ref{fig:Trap}. For values below the red dot-dashed curve, our assumption of free-streaming is valid. For larger $\epsilon$, our assumptions are invalidated and more intricate computations have to be done. Note, however, that such large-$\epsilon$ MCPs change the Sun's evolution even more drastically. Hence, our results tentatively hold for larger couplings as well.

\subsection{Analytical estimate} 

An analytical bound on MCPs from the Sun has been derived before in ref.~\cite{raffelt1999} by using a maximum excess luminosity for MCPs $L_{\text{MCP}}$ of $L_{\text{MCP}}/L_{\odot}<10\%$, thereby obtaining a limit $\epsilon<6\times 10^{-14}$ for $m_f=0 \ \text{eV}$. Interestingly, this bound is already very close to the one obtained by RGs and HB stars (both $\epsilon< 2\times 10^{-14}$~\cite{davidson2000}), where also an energy sink of $10\%$ of the star's created energy was assumed.

That the RGs and HB stars are not much better in constraining MCPs can be understood from the structure of the energy-loss rate  eq.~\eqref{eq:rate}. For $m_f=0$ the rate is proportional to $T^3 \omega_p^2\propto T^3 n_e$. Whereas the temperatures of the Sun, RGs and HB stars are not very different ($T_{\text{RG, HB}}/T_{\odot}\approx 8$), the densities 
vary by orders of magnitude $\omega_{p,\text{RG}}/\omega_{p,\odot}\approx 100$. While this difference leads to a higher emissivity of the denser stars, RGs and HB stars are more luminous than the Sun as well. This can be taken into account by comparing the energy creation rate per unit mass $\mathcal{Q}_\rho$. Dividing by the local density, $\mathcal{Q}_\rho$ becomes proportional to $T^3$ alone. Using a maximum energy loss for HB stars of $\mathcal{Q}_{\rho, \text{HB}}<10 \ \text{erg/s/g}$, an average temperature profile $\langle T_{\text{HB}}^3\rangle= 0.44$~\cite{Raffelt:1996wa}, and a maximum energy loss of the Sun of $\mathcal{Q}_{\rho,\odot} < 0.2 \ \text{
erg/s/g}$, we obtain that the bounds from~\cite{raffelt1999, 
davidson2000} behave like $\epsilon_{\odot}/\epsilon_{\text{HB}}\approx 3$, as seen above.

Conversely, a luminosity constraint for the Sun of $L_{\text{MCP}}<1\% \, L_{\odot}$, which corresponds to an order of magnitude improvement over the assumption by ref.~\cite{raffelt1999}, allows us to obtain similar bounds for RGs, HB stars and the Sun. We show in sec.~\ref{sec:MCPSun} that such an improvement can actually be obtained by carefully studying the SSMs.

\section{Statistical method}\label{sec:statistical}
 
We follow the statistical procedure that was introduced in ref.~\cite{villante2014} and applied in ref.~\cite{Vinyoles:2015aba} to obtain constraints on axions and massive hidden-photons. We construct a $\chi^2$-function that is used to quantify the quality of the solar models with respect to the 34 observables described in sec.~\ref{sec:Obs} using a method that is equivalent to a covariance matrix. To do so, we use the formula described in ref.~\cite{fogli2002}

\begin{equation}
\chi^2=\min_{\{\xi_I\}}\left[ \sum_Q \left( \frac{\delta_Q - \sum_I{\xi_I C_{Q,I}}}{U_Q}\right)^2 + \sum_I \xi_I^2 \right]\,,
\label{eq:xi}
\end{equation}
where $\delta_Q = 1 - \frac{Q_{\rm{model}}}{Q_{\rm{obs}}}$ is the relative difference between the model and observational data, $U_Q$ corresponds to the uncorrelated experimental errors and $C_{Q,I}$ are the  correlated theoretical uncertainties that are calculated by propagating the uncertainties of the input parameters $I$ of the SSMs to the observables, as in ref.~\cite{villante2014}. These input parameters are: age, diffusion coefficients, luminosity, opacity and six astrophysical S-factors for the most relevant nuclear reactions.

In this method, $\xi_I$ is a random variable with a standard normal distribution, and a shift $\xi_I C_{Q,I}$ is introduced to account for the correlations of the errors. This shift can be understood as a correction to the theoretical predictions of $Q_{\rm{model}}$ when the input parameters are varied. The penalty term $\sum_I \xi_I^2$ is included in the $\chi^2$ to account for the situation where the input parameters are away from their expected values, and thus, unrealistic in terms of the solar evolution. 
The values of $\xi_I$ that minimize the $\chi^2$ are called the pulls and they give information about tensions between parameters and data (e.g. it is not possible to change the luminosity and the opacity in different directions).
   
As we do not want our results to be affected by the solar abundance problem (see sec. \ref{sec:SSM}), we construct as a reference a new solar model by letting the composition free, within realistic input parameters, in order to minimize the $\chi^2$-function and to best reproduce the observations.  Following ref.~\cite{villante2014}, we group the elements into two different classes, \textit{refractories} (Mg, S, Si, Fe) and \textit{volatiles}  (C, N, O, Ne). We vary them with a multiplicative factor $(1+\delta_z)$ and look for the values of $\delta_{z,\rm{vol}}$ and $\delta_{z,\rm{ref}}$ that minimize the $\chi^2$ function. For the case without dark emission, the solar model thus obtained is called the \textit{best-fit model}~\cite{villante2014, Vinyoles:2015aba}.

For the case with MCP emission, we proceed in a similar way. We fix an MCP mass $m_f$ and then, for different minicharges $\epsilon$, calculate the $\chi^2$-function letting the composition free as before. Comparing the resulting $\chi^2$-values, we can obtain a bound in the ($\epsilon$, $m_f$) plane as is shown in sec.~\ref{sec:results}. We stress that although we vary $\epsilon$ while fixing $m_f$, this corresponds to a $\chi^2$ with two degrees of freedom.

 As has been pointed out in ref.~\cite{Vinyoles:2015aba}, this procedure is valid when the emission depends only on the stratification of the temperature and mass density in the solar interior and not on its detailed composition. This is the case for MCPs as can be seen from the analytical emission rates in eqs.~\eqref{eq:rate} and~\eqref{eq:rate2}. Therefore, our results will not depend on the composition, and we present all results using the more recent composition compilation AGSS09.
 
\section{Minicharged particles in the Sun}\label{sec:MCPSun}

\begin{figure}\centering
 \includegraphics[width=0.49\textwidth]{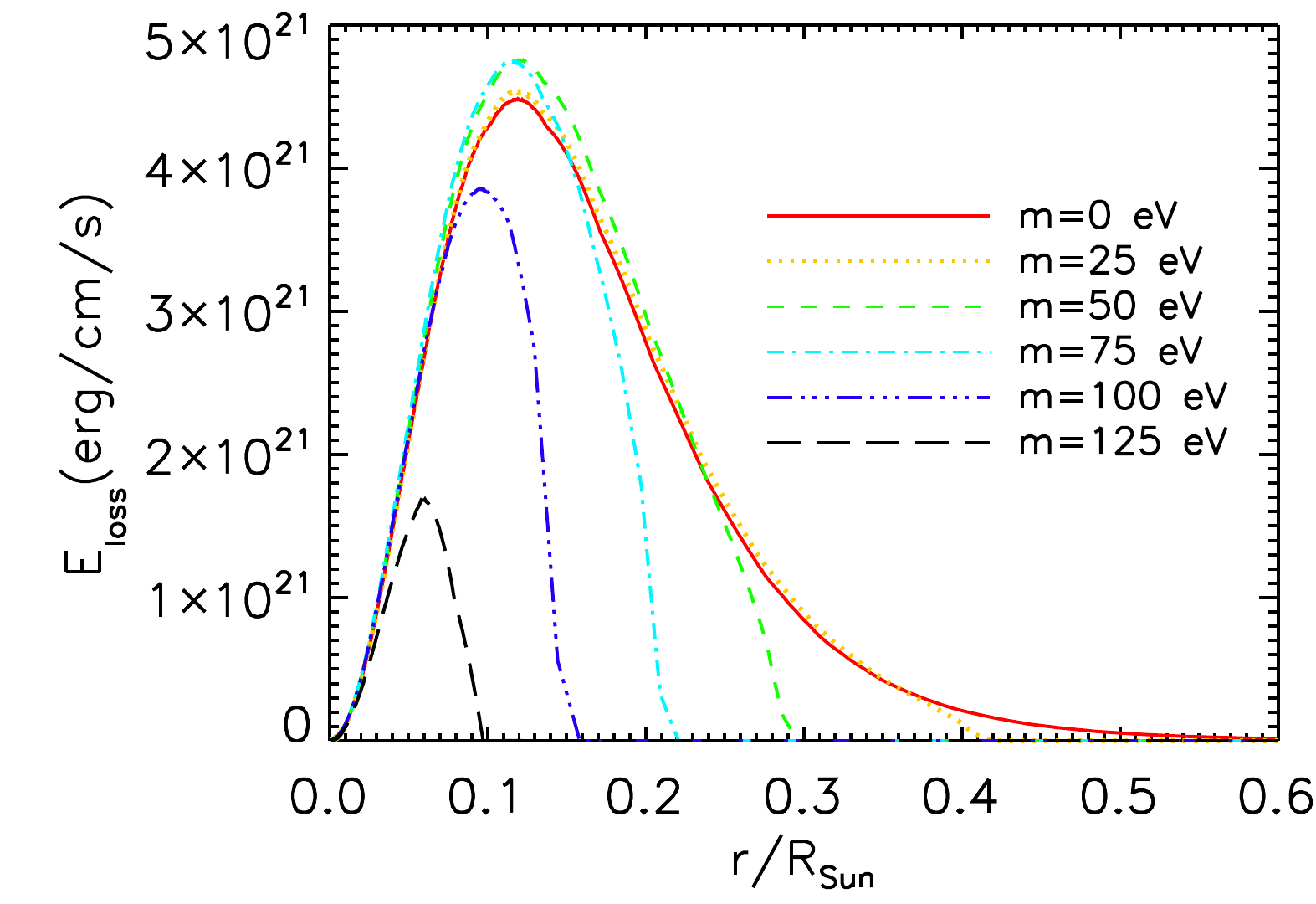}
 \includegraphics[width=0.49\textwidth]{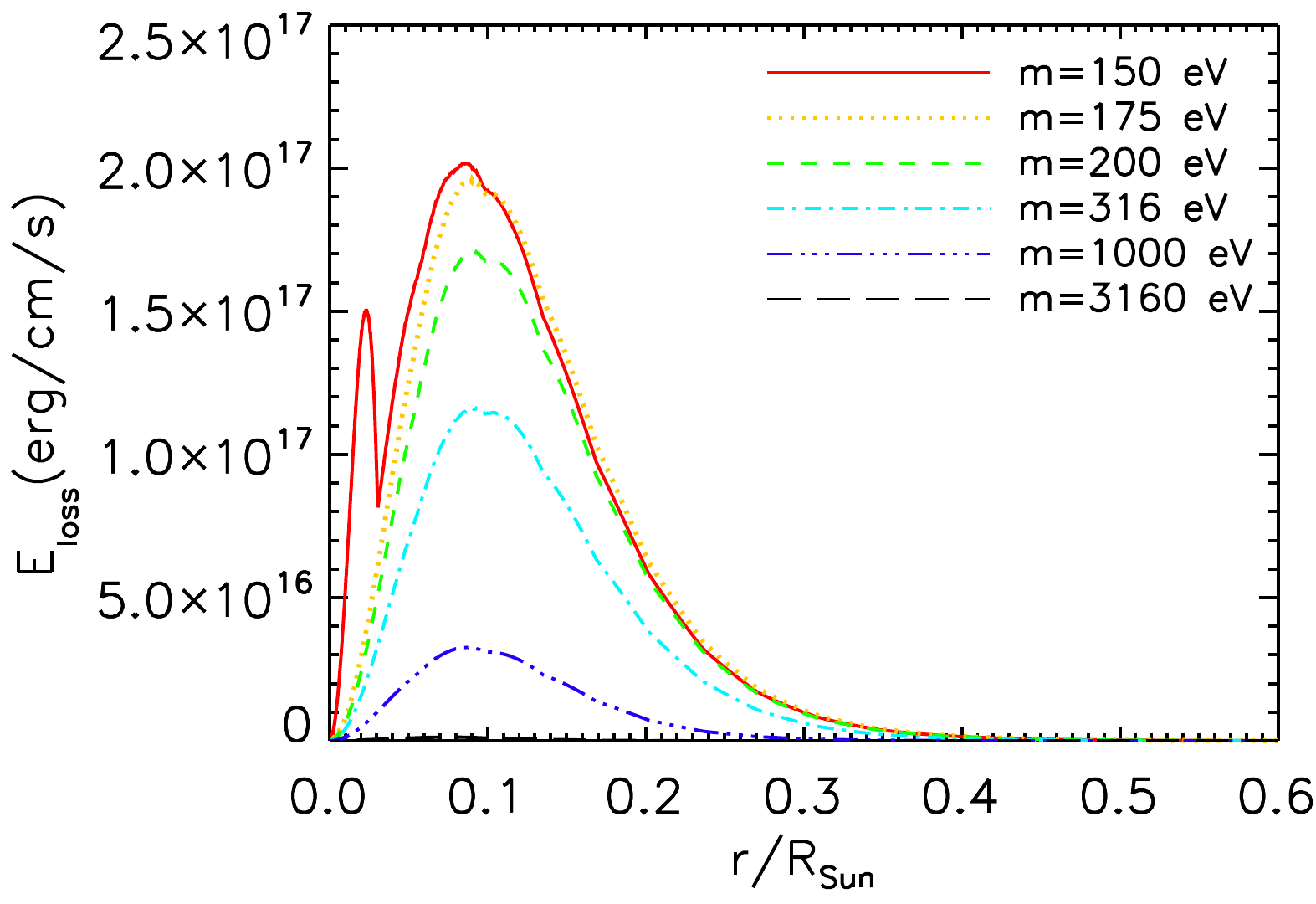}
 \caption{Energy-loss distribution of spheres with radius $r$ for different values of $m_f$. The left panel corresponds to the case where $2 m_f < \omega_{p}$  so that the emission is not suppressed at least somewhere inside the Sun. The right panel corresponds to the suppressed emission $2 m_f > \omega_{p}$.}
 \label{fig:mcpemission}
 \end{figure}
\begin{figure}\centering
 \includegraphics[width=0.49\textwidth]{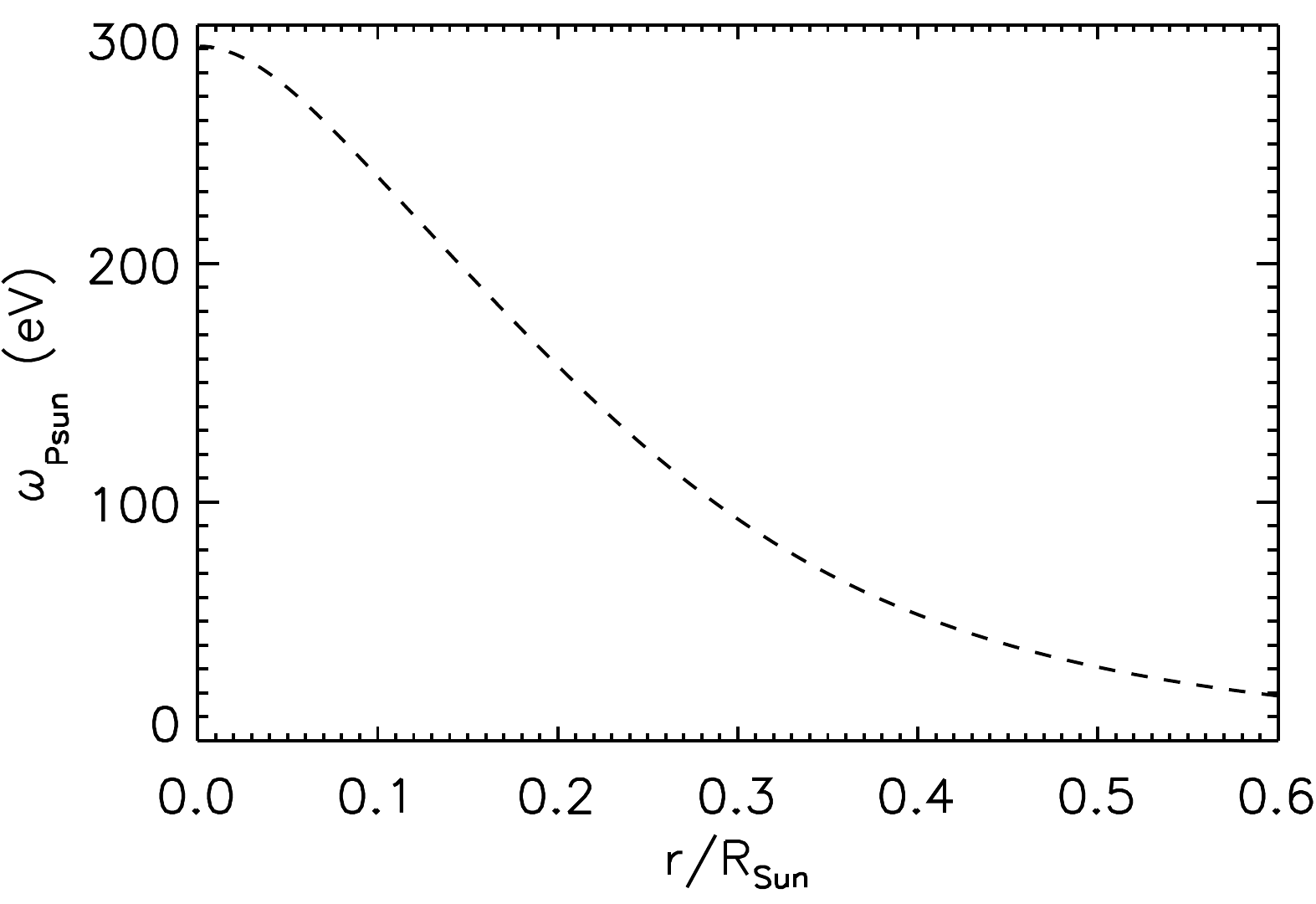}
 \caption{Solar plasma frequency as a function of the solar radius for the SSM ($\epsilon = 0$).}
 \label{fig:wp}
 \end{figure}

We now include the energy loss of MCPs in the SSMs. The MCPs change the internal structure of the Sun, especially the temperature and density profile. In fig.~\ref{fig:mcpemission}, we present the exotic energy-loss distribution as a function of the solar radius for different $m_f$ with $\epsilon = 2.2 \times 10^{-14}$ and in fig.~\ref{fig:wp} we show the values of $\omega_p$ as a function of the solar radius. For $m_f=0$, on-shell plasmon decay is possible everywhere inside the Sun because $2m_f<\omega_p$ is always fulfilled. For $m_f>0$, less dense regions do not allow for on-shell decay anymore and the weaker off-shell decay occurs. Hence, the emission rate is strongly suppressed for larger radii. 

 On-shell emission is completely suppressed when $2 m_f > \omega_p$ everywhere inside the Sun. The maximum value for the plasma frequency of the SSMs, we consider in the following, peaks at around $\omega_p\sim 290\ \text{eV}$ such that for $m_f\geq 175 \ \text{eV}$ only off-shell decay is possible.  For  $m_f=150 \ \text{eV}$, fig.~\ref{fig:mcpemission} shows a two-peaked structure. The reason for this is that quasi on-shell emission occurs in a small sphere around the core of the Sun. Here, the emission rate is dominated by the flank of the quasi-particle Breit-Wigner profile which rises quickly when $\omega_p$ approaches $2m_f$.

 As apparent from fig.~\ref{fig:mcpemission},  different values of $m_f$ lead to different energy-loss distributions and, thus, different results for the solar structure. To show the different effects, in figs.~\ref{fig:soundspeed} and~\ref{fig:neutrino} we have plotted the resulting sound speed profiles and neutrino fluxes for MCP models with four different masses. The results are the outputs of AGSS09 when including MCPs and before the variation of the composition. The results for $Y_s$ and $R_{\rm{CZ}}$ are not presented in this paper because they do not change substantially for the range of $\epsilon$ considered here. They will contribute to the global value of the $\chi^2$ but they are not significant in giving constraints on $\epsilon$.

\subsection{Sound speed profiles} 

\begin{figure}\centering
 \includegraphics[width=0.4\textwidth]{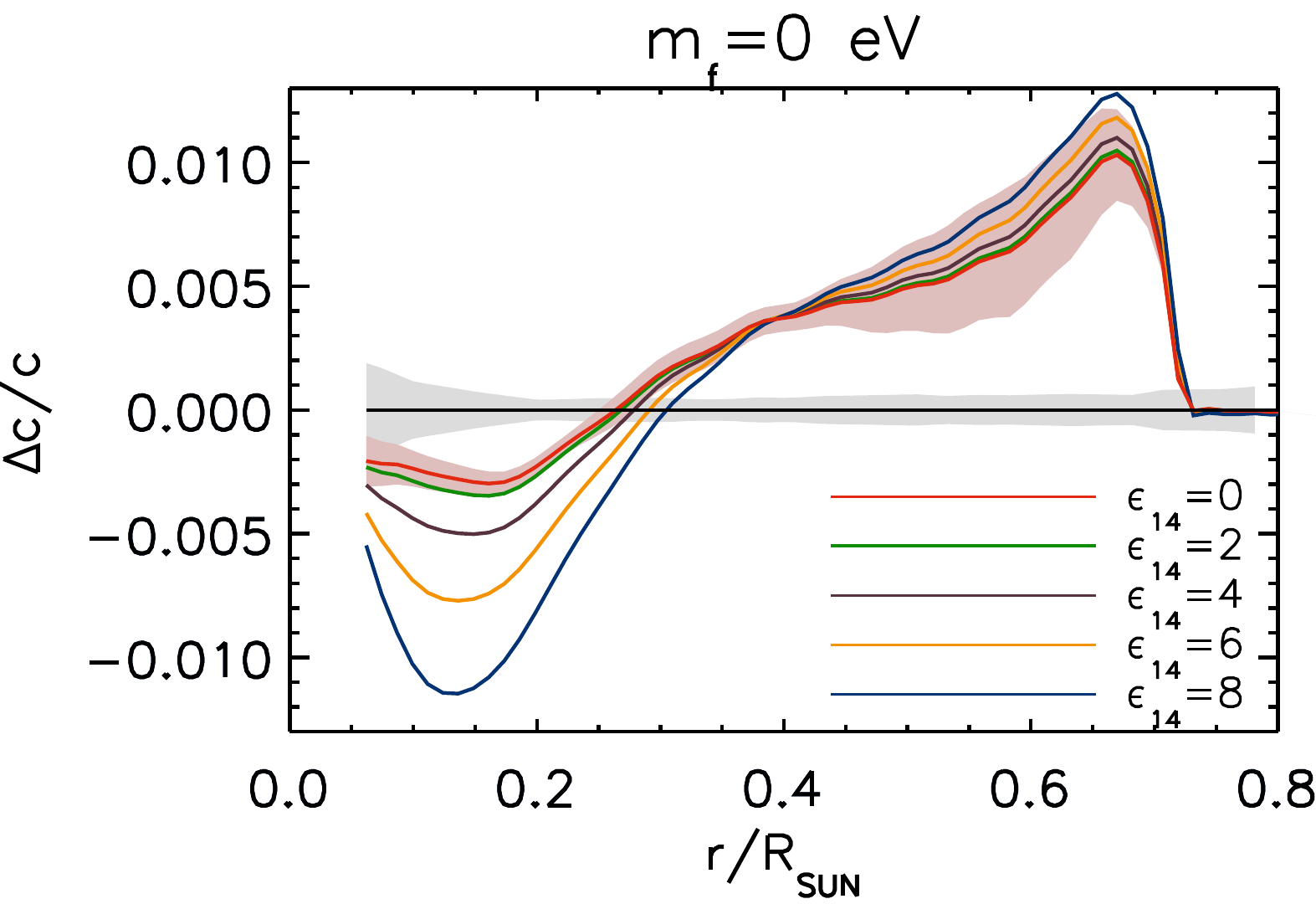}
  \includegraphics[width=0.4\textwidth]{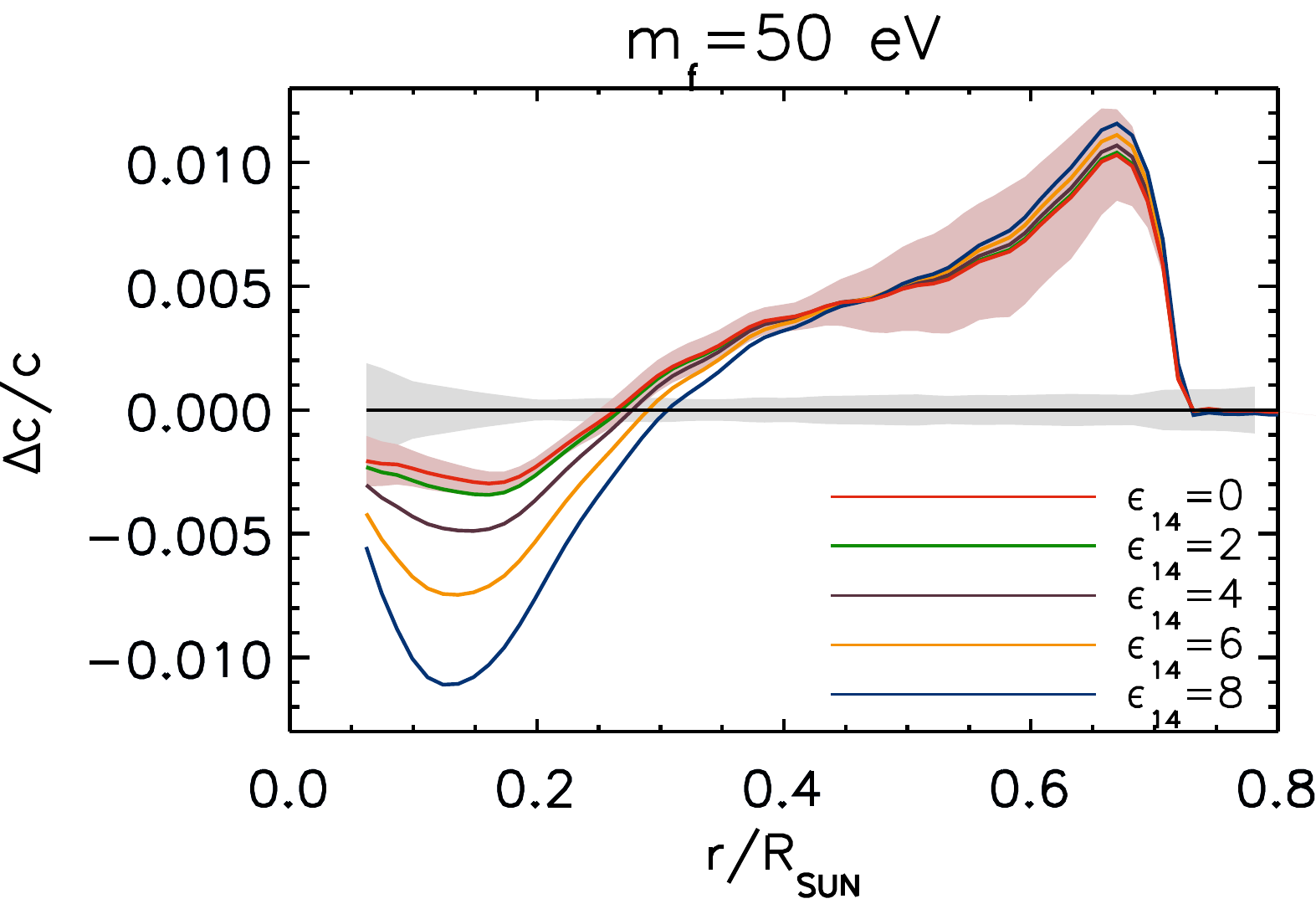}\\
  \includegraphics[width=0.4\textwidth]{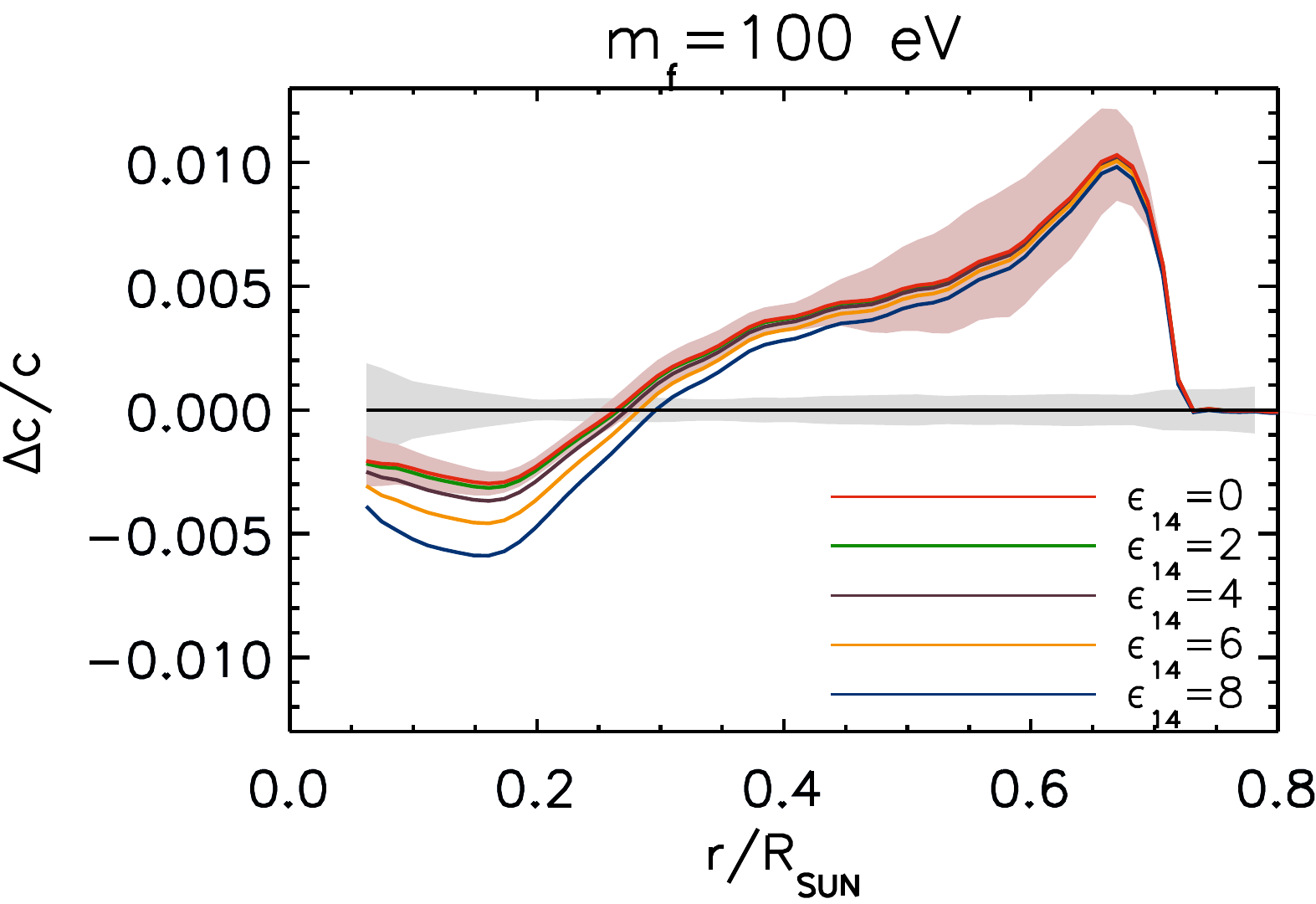}
 \includegraphics[width=0.4\textwidth]{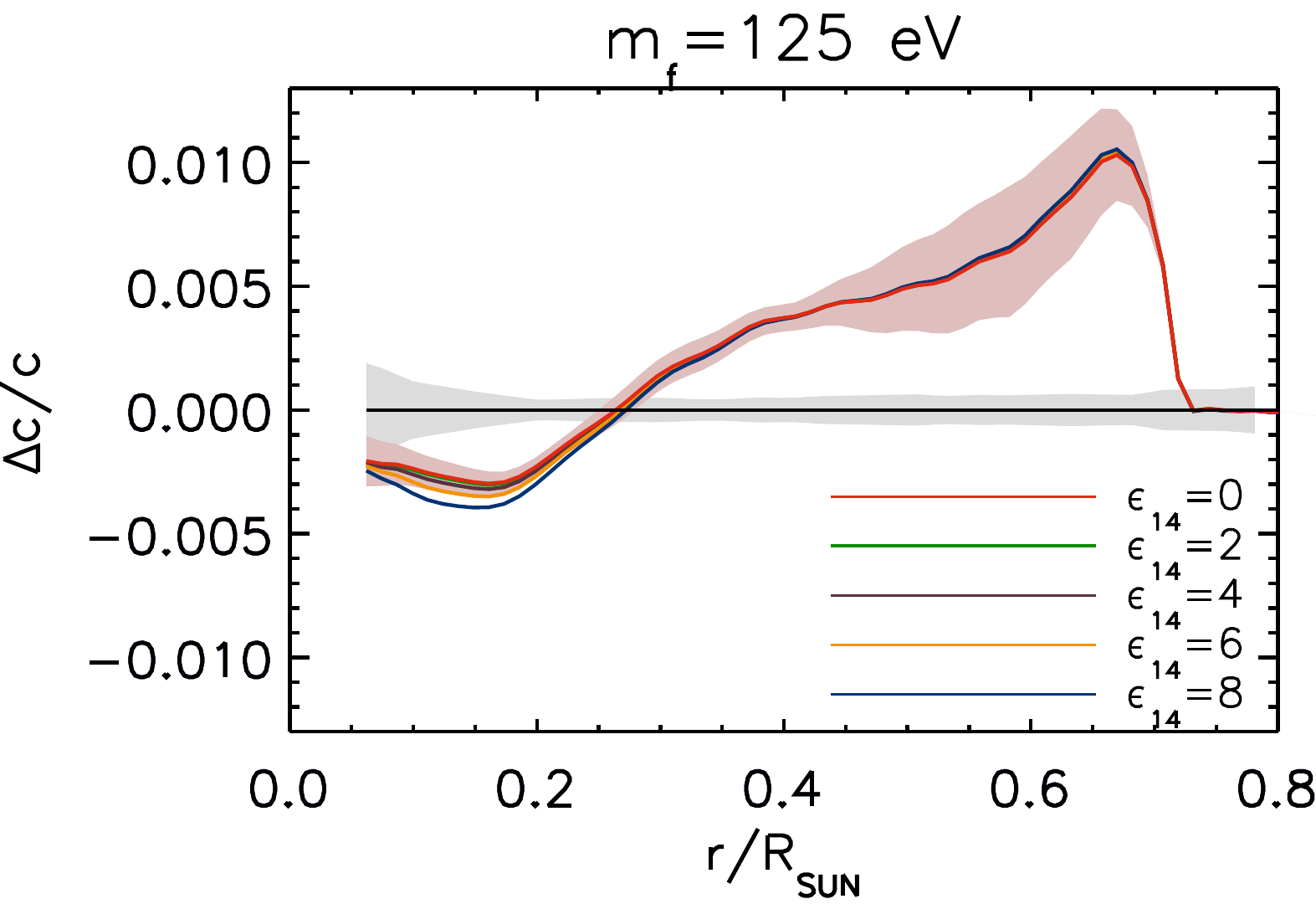}
 \caption{Sound speed profiles of AGSS09. Each of the panels represents one value for $m_f$. Grey shade corresponds to the observational errors and red shade to the theoretical ones. }
 \label{fig:soundspeed}
 \end{figure}

The effect of the radial distribution of the energy-loss on the sound speed profile can be understood by comparing MCP models with equal minicharges but different masses. For small values of $m_f$, the deviation of the profile with energy-loss from the Standard Solar Model without extra energy-loss is apparent all along the sound speed profile (see~fig.~\ref{fig:soundspeed}). For larger masses $m_f$, the total energy-loss is reduced so that the change in the sound speed profile becomes less relevant. Moreover, the MCP emission is more localized in the inner part of the Sun where on-shell emission still takes place, so that the sound speed profile is altered more drastically in the inner region.

\subsection{Neutrino fluxes}

\begin{figure}\centering
  	\includegraphics[width=0.35\textwidth]{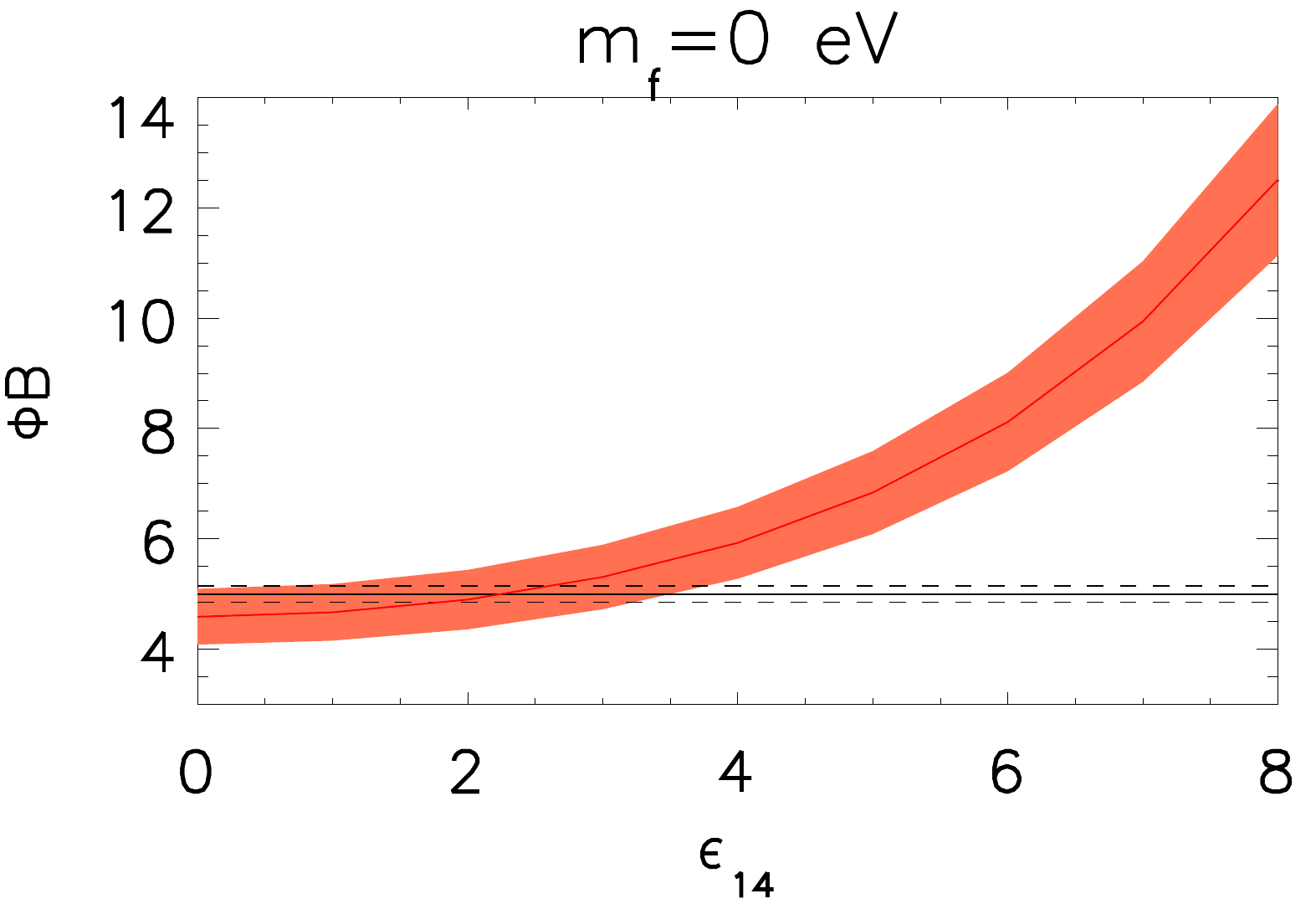}
   	\includegraphics[width=0.35\textwidth]{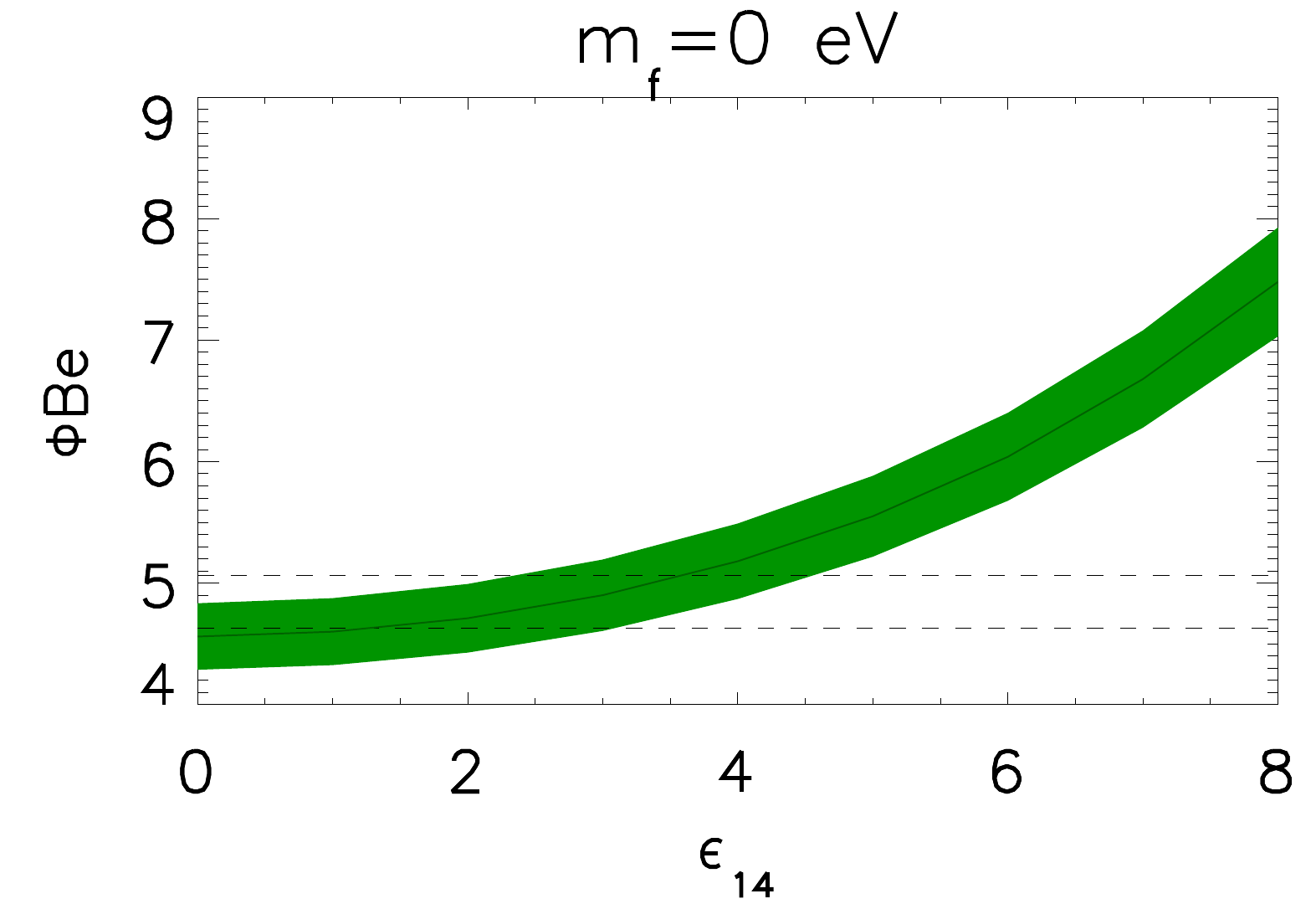}\\
    \includegraphics[width=0.35\textwidth]{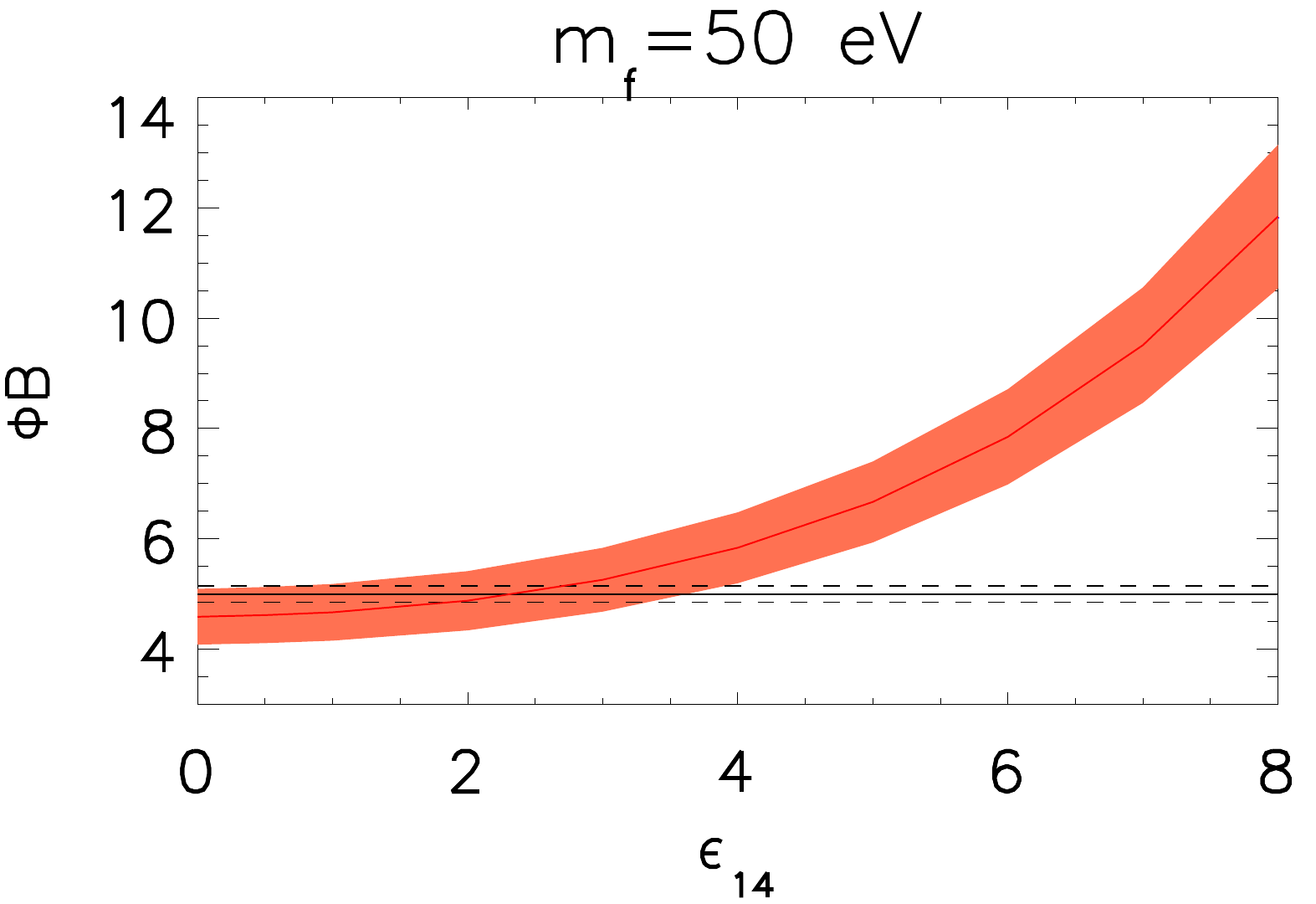}
    \includegraphics[width=0.35\textwidth]{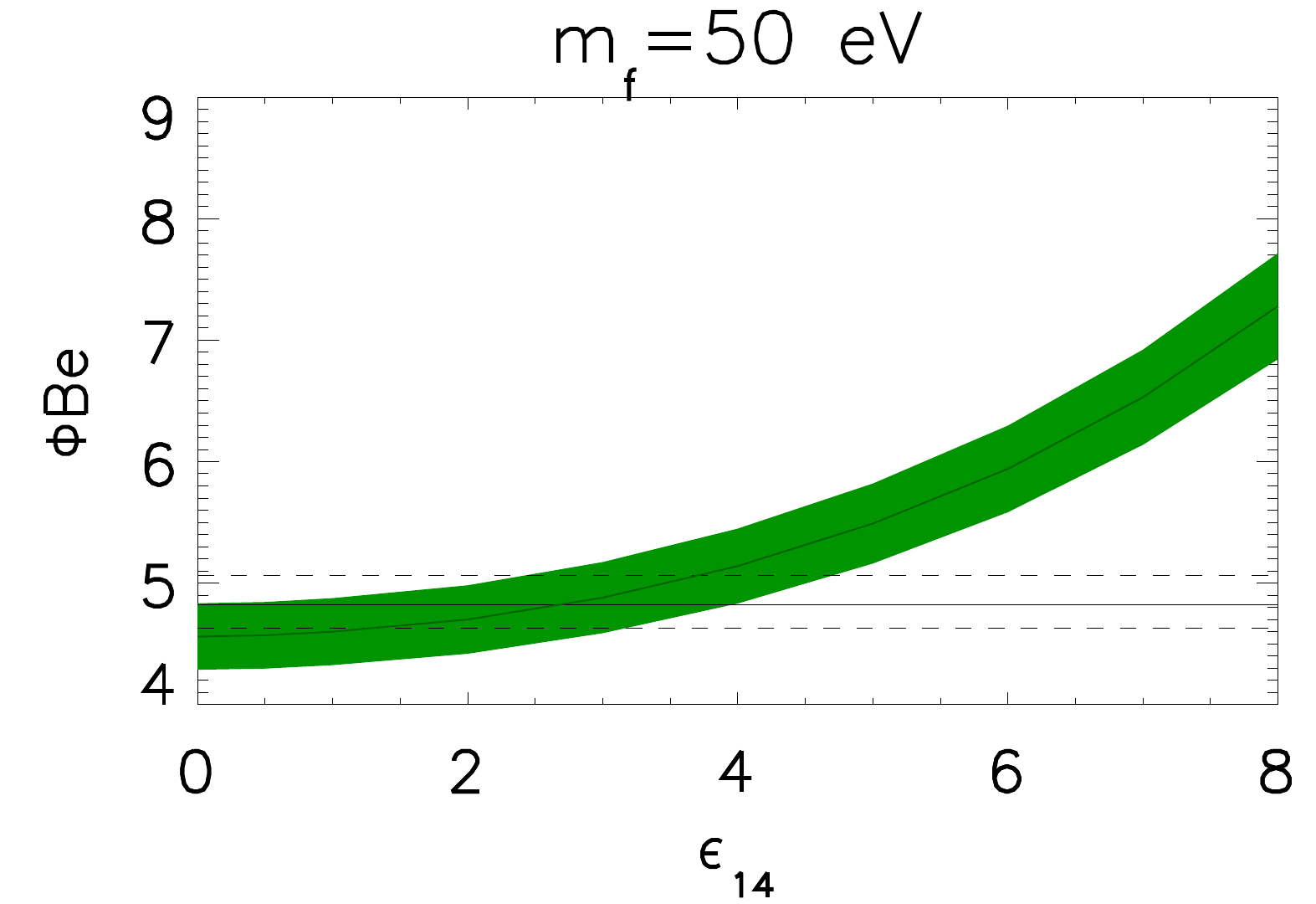}\\
  \includegraphics[width=0.35\textwidth]{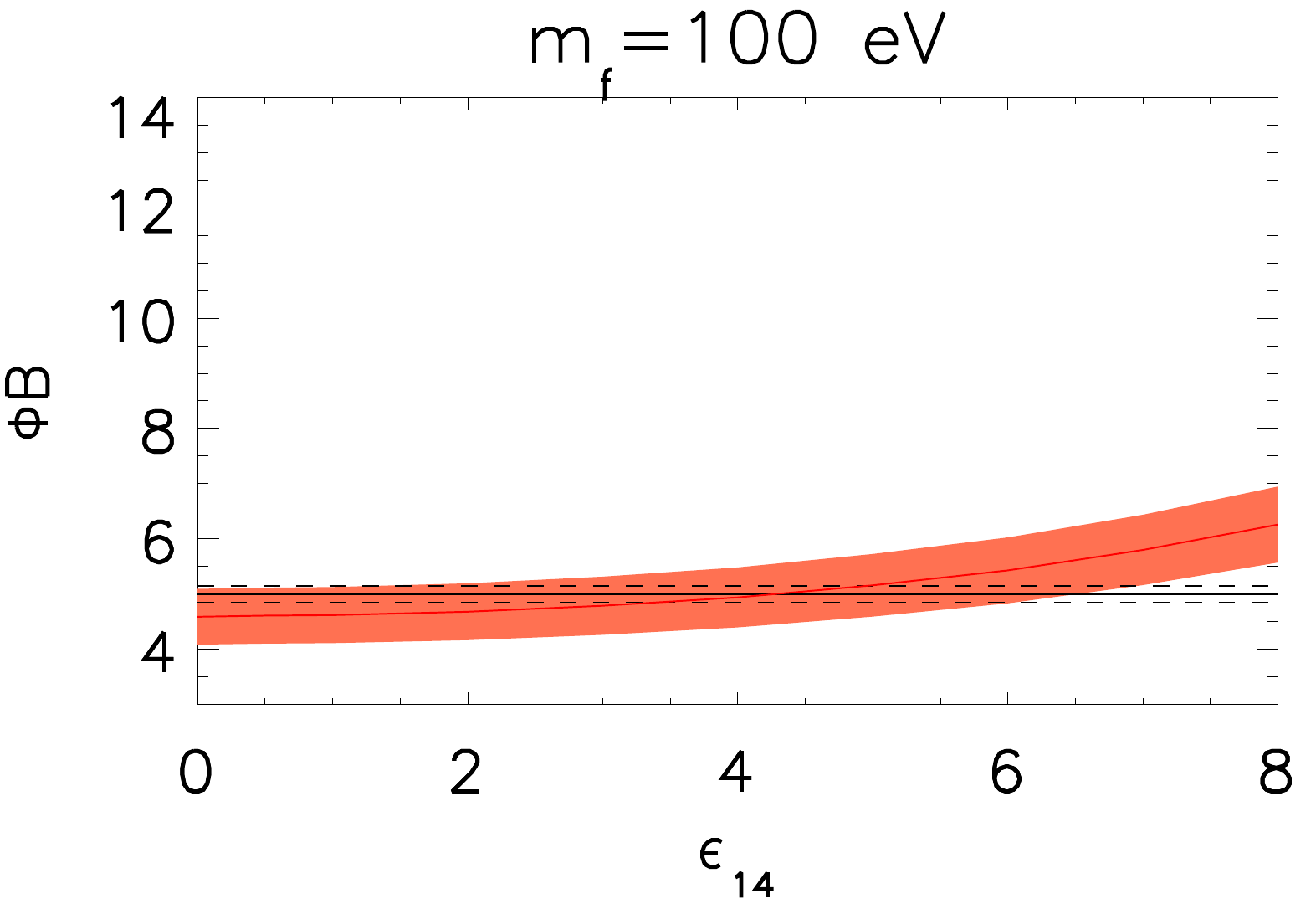}
   \includegraphics[width=0.35\textwidth]{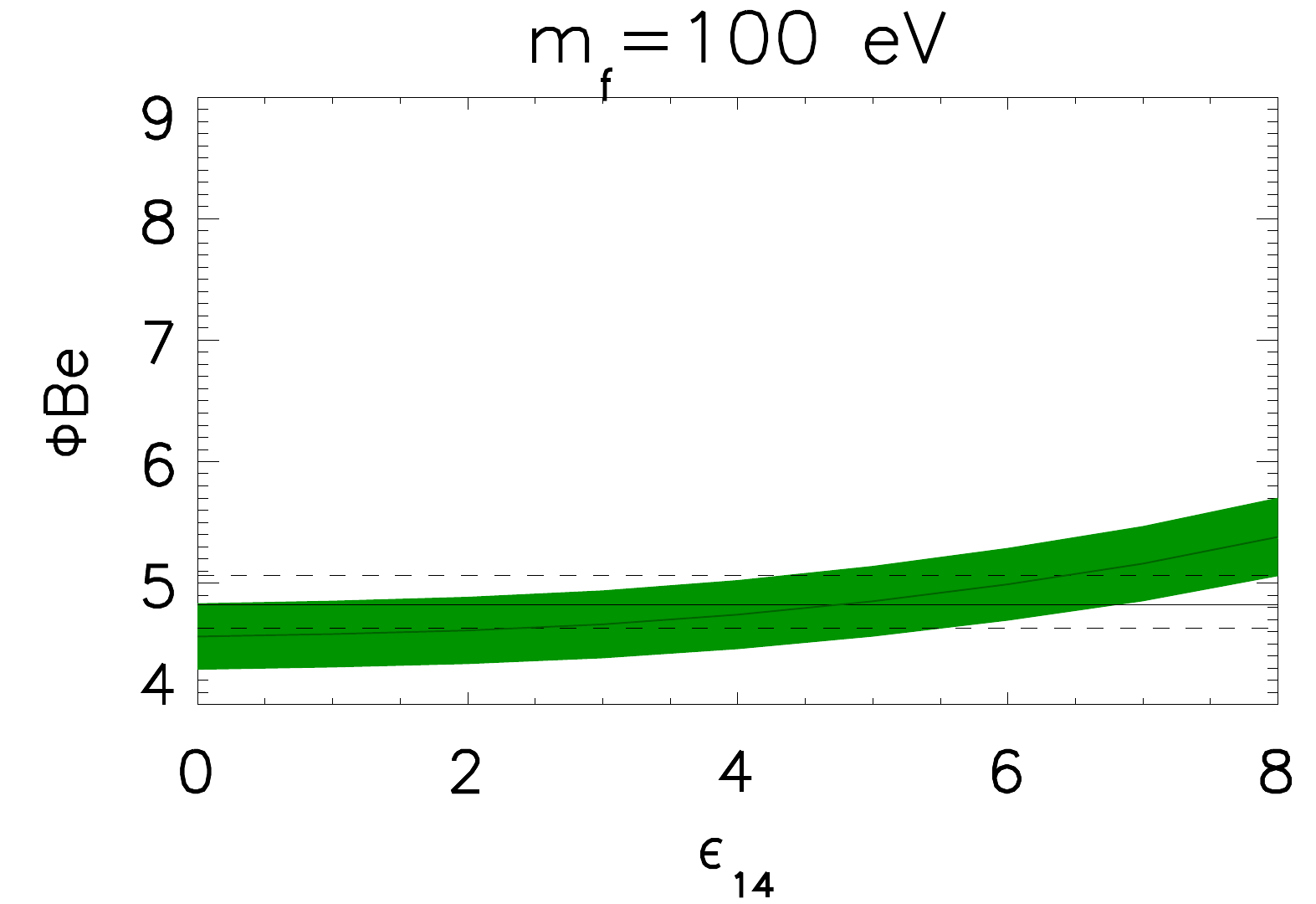} \\
   \includegraphics[width=0.35\textwidth]{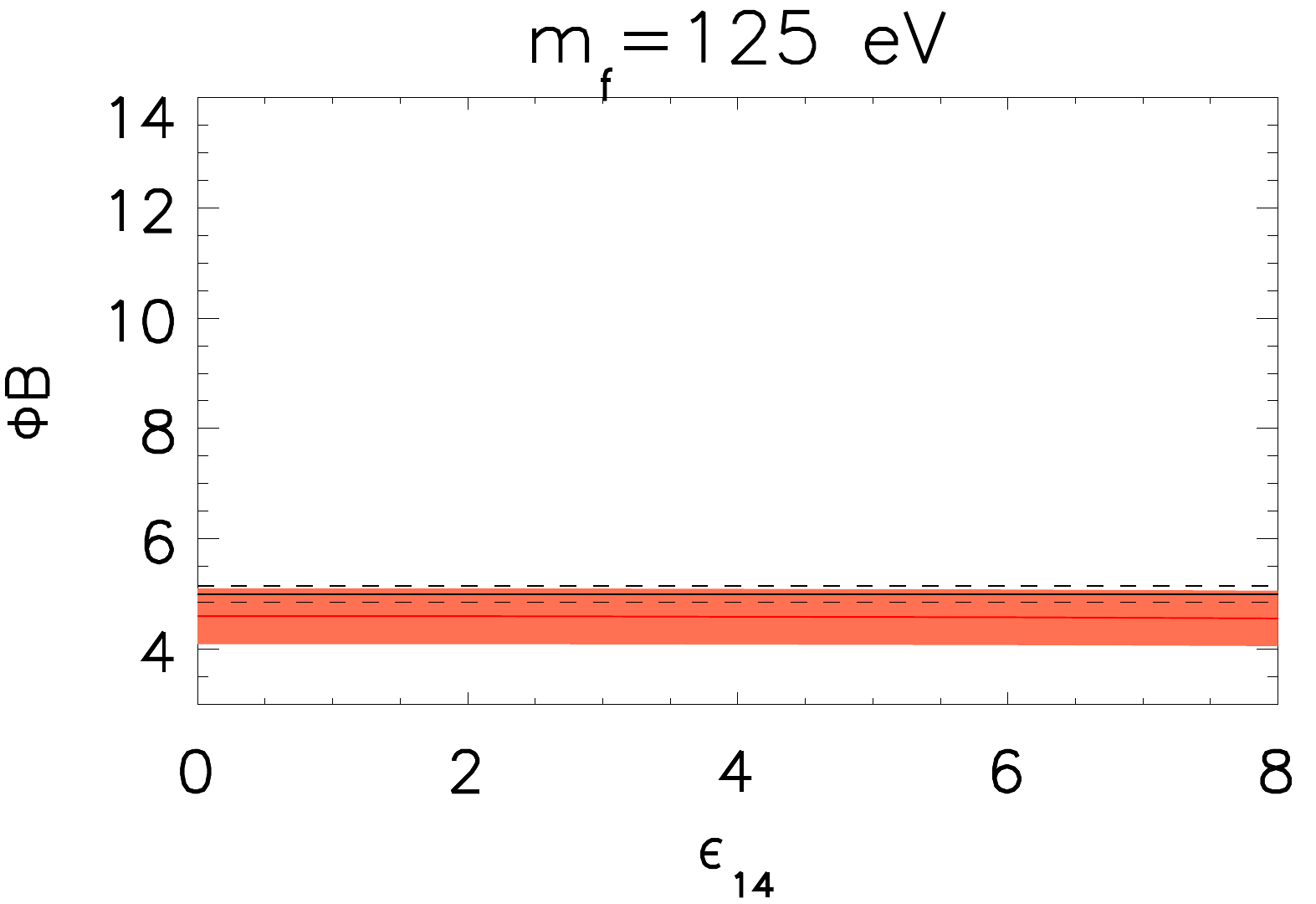}
    \includegraphics[width=0.35\textwidth]{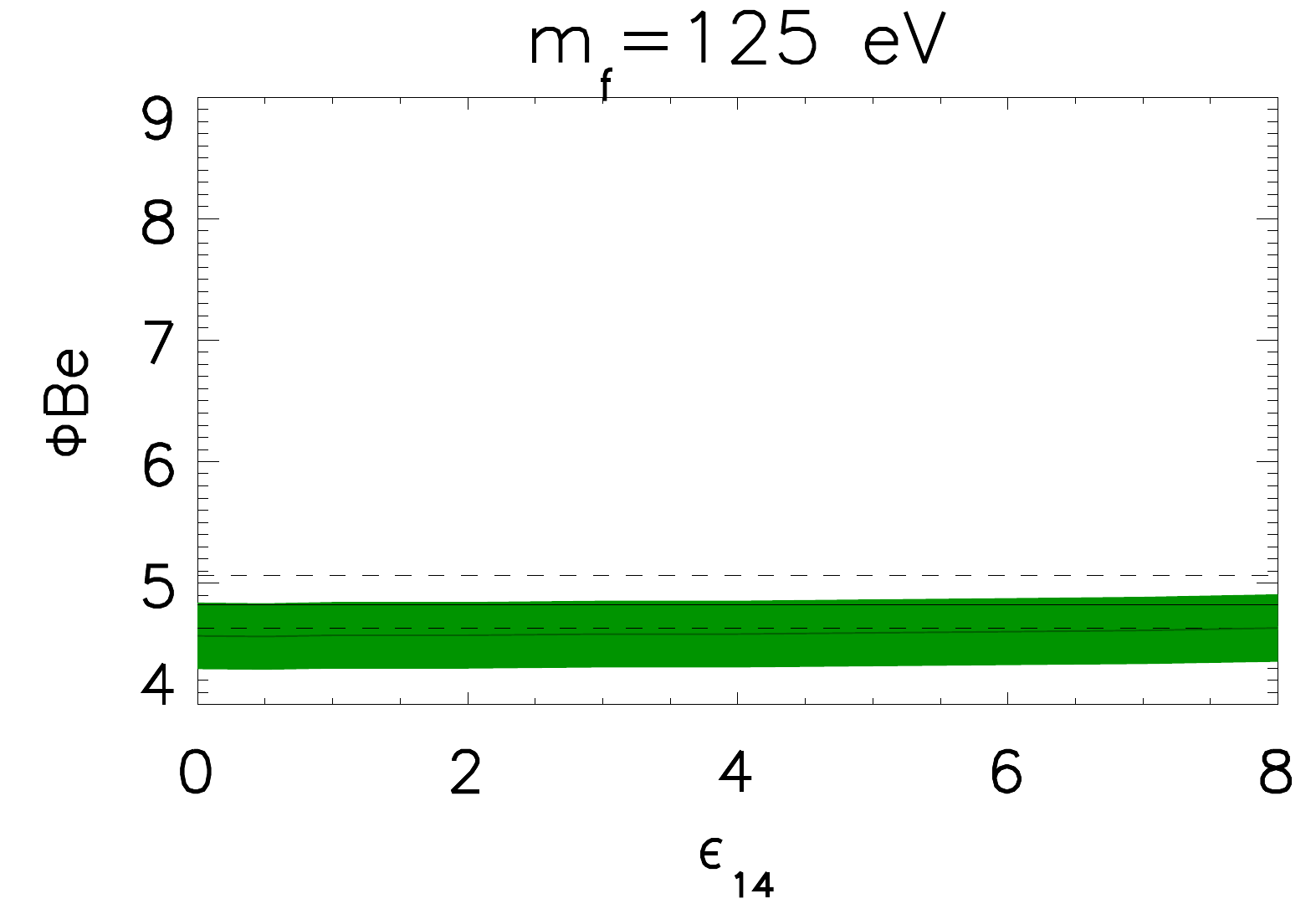} \\
 \caption{$\rm{^7Be}$ (right panel) and $\rm{^8B}$ (left panel) neutrino fluxes as a function of $\epsilon_{14} = \epsilon \times 10^{14}$. Each row represents one value for $m_f$. Dashed lines correspond to the observational errors whereas the colored regions correspond to theoretical errors. The fluxes are in $10^9\, \rm{cm^{-2} s^{-1}}$ for $\rm{\Phi(^7Be)}$ and $10^6 \,\rm{cm^{-2} s^{-1}}$ for $\rm{\Phi(^8B)}$.}
 \label{fig:neutrino}
 \end{figure}

When we calculate solar models with exotic energy-loss, we observe changes on the neutrino fluxes as can be seen in fig.~\ref{fig:neutrino}. These changes in the neutrino fluxes can be understood in the following way: When we add some extra energy-loss to a solar model, we need to increase the energy production through nuclear reactions in order to reach the observed solar luminosity ($\rm{L_\odot = L_{nuc} - L_{MCP}}$) resulting in an increase of the neutrino fluxes.

This situation should not be confused with the one described in ref.~\cite{bahcall1996} (a case without exotic energy losses) where it is  shown that the \textit{pp} neutrino flux depends on a negative power of the temperature. The difference is that in the situation of ref.~\cite{bahcall1996}, the condition to fulfill is $\rm{L_\odot = L_{nuc}}$, so that when neutrinos are neglected all the energy created through nuclear reactions contributes to the solar luminosity. 
Then, if we increase the temperature, the ratio between the ppI process of the pp-chain and the more energetic ppII process will change favoring the latter. Hence, the \textit{pp} production decreases to maintain $\rm{L_\odot = L_{nuc}}$. In our work, we fulfill $\rm{L_\odot + L_{MCP} = L_{nuc}}$ instead so that the nuclear energy has to increase in order to account for the extra energy loss.

The reaction rates of $\rm{^7Be + e^- \rightarrow {}^7Li + \nu_e}$ and $\rm{^8 B \rightarrow {}^7Be + e^+ + \nu_e}$ increase strongly with rising temperature making these neutrino fluxes very sensitive to temperature changes. Their emissivity of high energy neutrinos can be measured experimentally and poses an independent probe of the solar interior.

In fig.~\ref{fig:neutrino}, we present the $\rm{^8B}$ (left panel) and $\rm{^7Be}$ (right panel) neutrino fluxes as a function of $\epsilon$ for models with different masses. For the cases with $m_f = 0,50,100 \ \text{eV}$, both neutrino fluxes increase with increasing $\epsilon$ and, hence, energy-loss. This shows that in the regions where those nuclear reactions take place the temperature is higher than the SSM.

\begin{figure}[t]\centering
  	\includegraphics[width=0.49\textwidth]{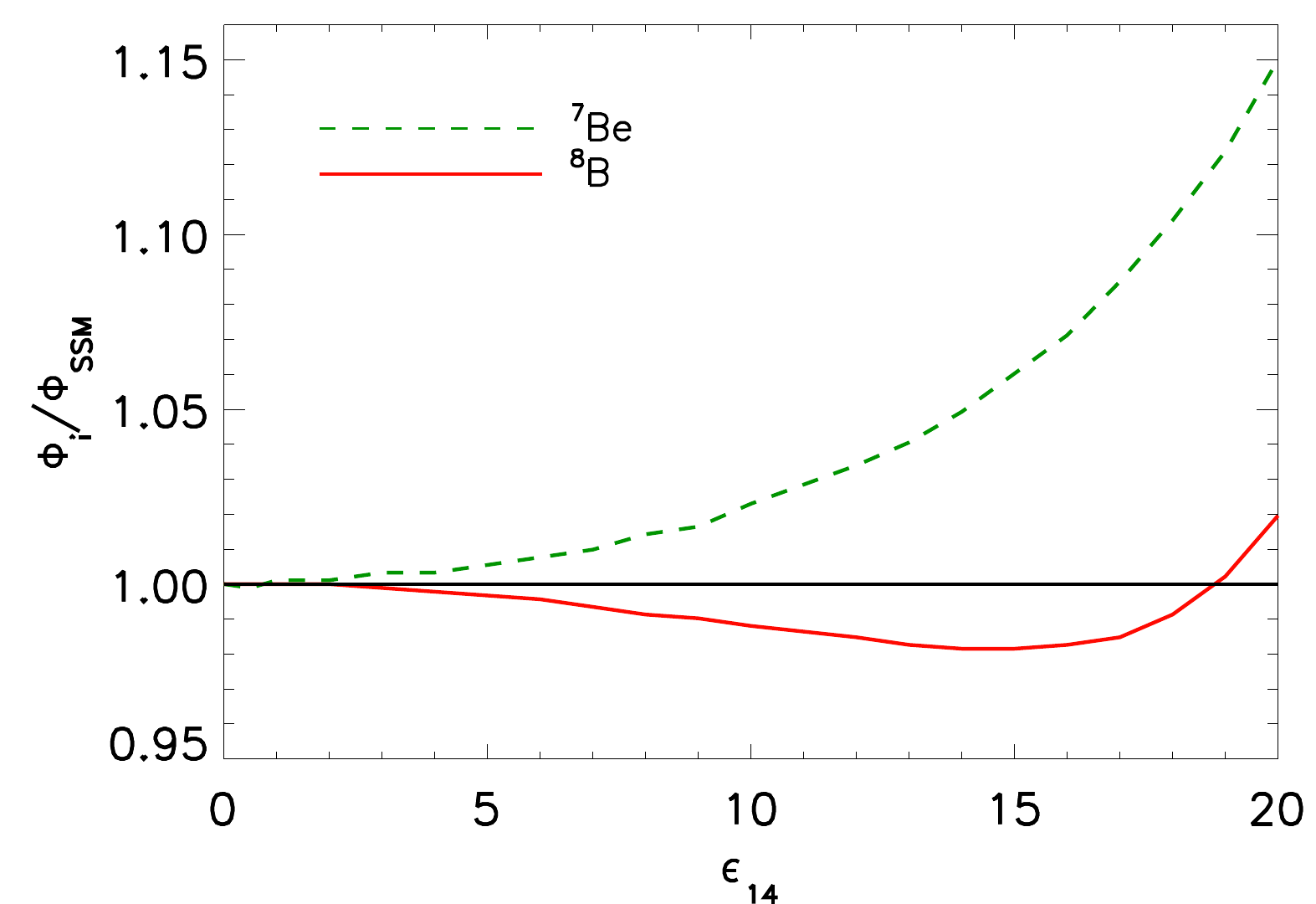}
  	\includegraphics[width=0.49\textwidth]{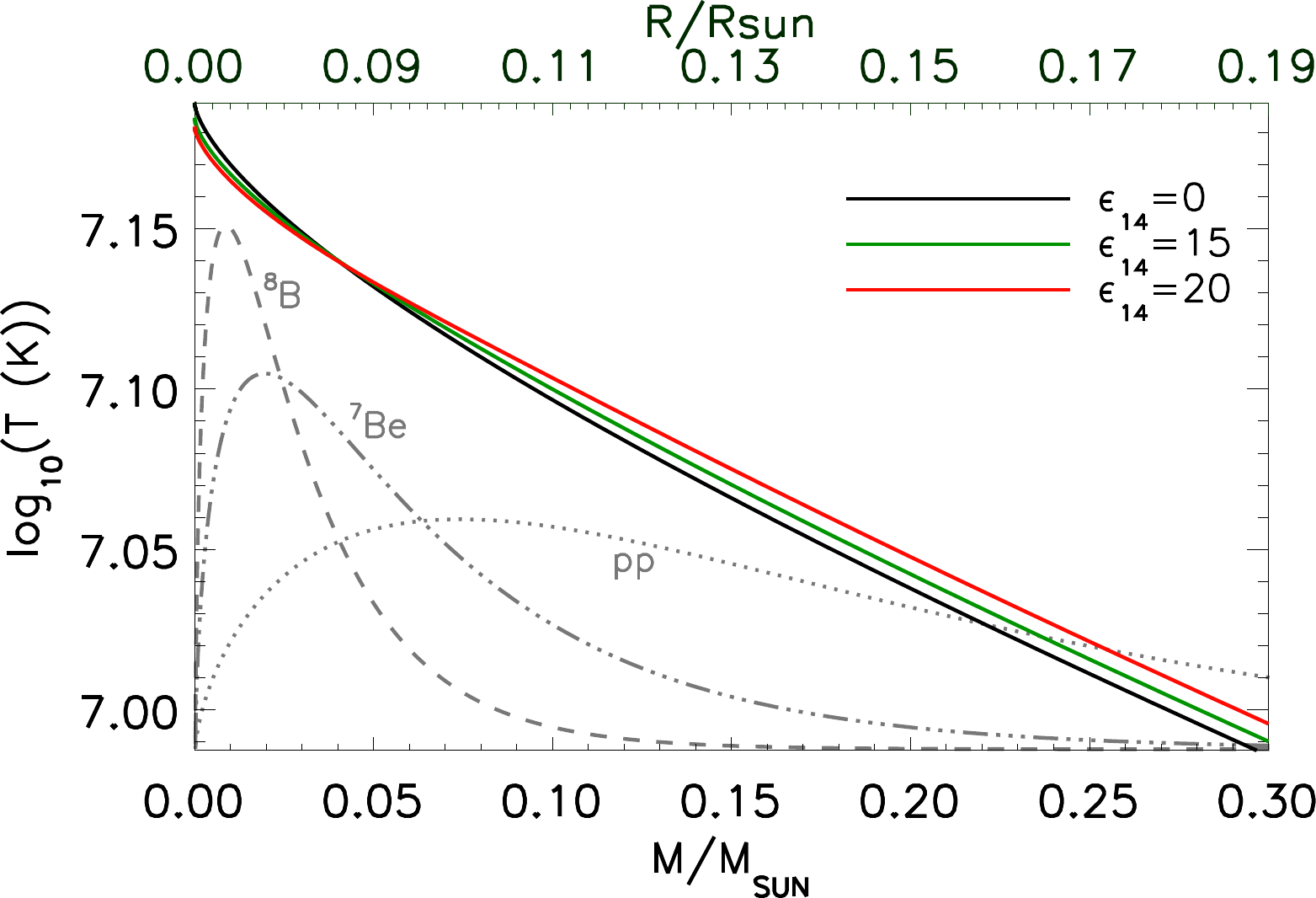}
  	\caption{\small{\emph{Left panel:} Relative neutrino flux of the $m_f=125 \ \text{eV}$ model with respect to AGSS09 SSM as a function of $\epsilon_{14}$. The green dotted line corresponds to the $^7\rm{Be}$ neutrino flux and the solid red line to the $^8\rm{B}$ neutrino flux. \emph{Right panel:} Temperature profile for the model with $m_f=125 \, \text{eV}$ for three different values of $\epsilon$ as a function of mass and radius. The normalized probability distribution of the production as a function of radius for the different neutrino fluxes are given by grey dotted, dashed and dash-dotted lines.}}
 \label{fig:m125}
 \end{figure}

The case $m_f = 125 \ \rm{eV}$ is rather peculiar. In fig.~\ref{fig:neutrino}, we observe minimal changes in the $\epsilon$-range that we have plotted, meaning that the energy-loss rate is too low  to produce noticeable changes in the Sun. In fig.~\ref{fig:m125} (left panel), we extend the parameter space and show the relative variation  in the range of $\epsilon$ where the changes are significant. We can observe that the trend of the $^8\rm{B}$ flux differs from the cases seen before. The $^8 \rm{B}$ flux initially decreases with $\epsilon$ while the flux of $^7 \rm{Be}$ increases with it. 

This interesting behavior can be explained if we study how the temperature profile in the interior behaves for cases with different $\epsilon$. In the right panel of fig.~\ref{fig:m125}, we show this profile for three different cases with $\epsilon=0$, $\epsilon=15 \times 10^{-14}$ ($\sim 0.1~\rm{L_\odot}$) and $\epsilon=20 \times 10^{-14}$ ($\sim 0.2~\rm{L_\odot}$). In the interior ($r < 0.08~\rm{R\odot}$), the temperature decreases when we include extra energy-losses,
while in the rest of the Sun the temperature increases. The change in the temperature profile results in different flux variations depending on where the reactions take place. In the right panel of fig.~\ref{fig:m125}, we also plotted the production probability distribution as a function of radius of the fluxes $\rm{\Phi(pp)}$, $\rm{\Phi(^8B)}$ and $\rm{\Phi(^7Be)}$. We see that $\rm{\Phi(^8B)}$ is localized in the region where the temperature decreases with increasing energy-loss, which explains why the flux initially decreases with $\epsilon$. On the other hand, the  $\rm{\Phi(^7 Be)}$ flux lies in both (increasing and decreasing temperature) regions. The result is going to be a combination of the decrease of the $^7 \rm{Be}$ production rate in the inner part and the increase of the production of neutrinos in the more exterior 
part. In this particular situation, this results in an increase of the $^7 \rm{Be}$ neutrino production as a function of $\epsilon$. In general, it is hard to predict the behavior of the neutrino fluxes for highly localized exotic emission since the variation depends on more factors than just the temperature profile (abundance of protons, $^3\rm{He}$ equilibrium abundance, etc.).

 Finally, we would like to mention that some works, as e.g. ref.~\cite{bahcall1996}, show that the neutrino fluxes can be expressed as power-laws of the central temperature of the Sun ($\Phi_i \propto \rm{T_c^m}$). In contrast, in this work, we have shown that this relation should be used with great care because it does not apply to all situations. We show that one should take the whole temperature profile into consideration and not just focus on the central temperature of the Sun.

\section{Results}
\label{sec:results}

We have calculated the $\chi^2$-function for different MCP masses varying the composition as described in sec~\ref{sec:statistical}. For the SSM, we obtain $\chi^2=38.5$ using 34 observables, which is also the minimal $\chi^2$ for almost all the masses. For some values of $m_f$, however, the minimum is not exactly at $\epsilon=0$ (SSM) but close to it, which
could be due to some small tensions between the sound speed profile and the radius at the convective zone. However, these deviations are not statistically significant. Hence, we come to the conclusion that for all the models, the presence of MCPs does not improve the agreement of the models with the observations. On the other hand, when the energy-loss starts being important, $\chi^2$ grows smoothly and we use this function to limit an MCP contribution.

\begin{figure}[t]\centering
  	\includegraphics[width=0.49\textwidth]{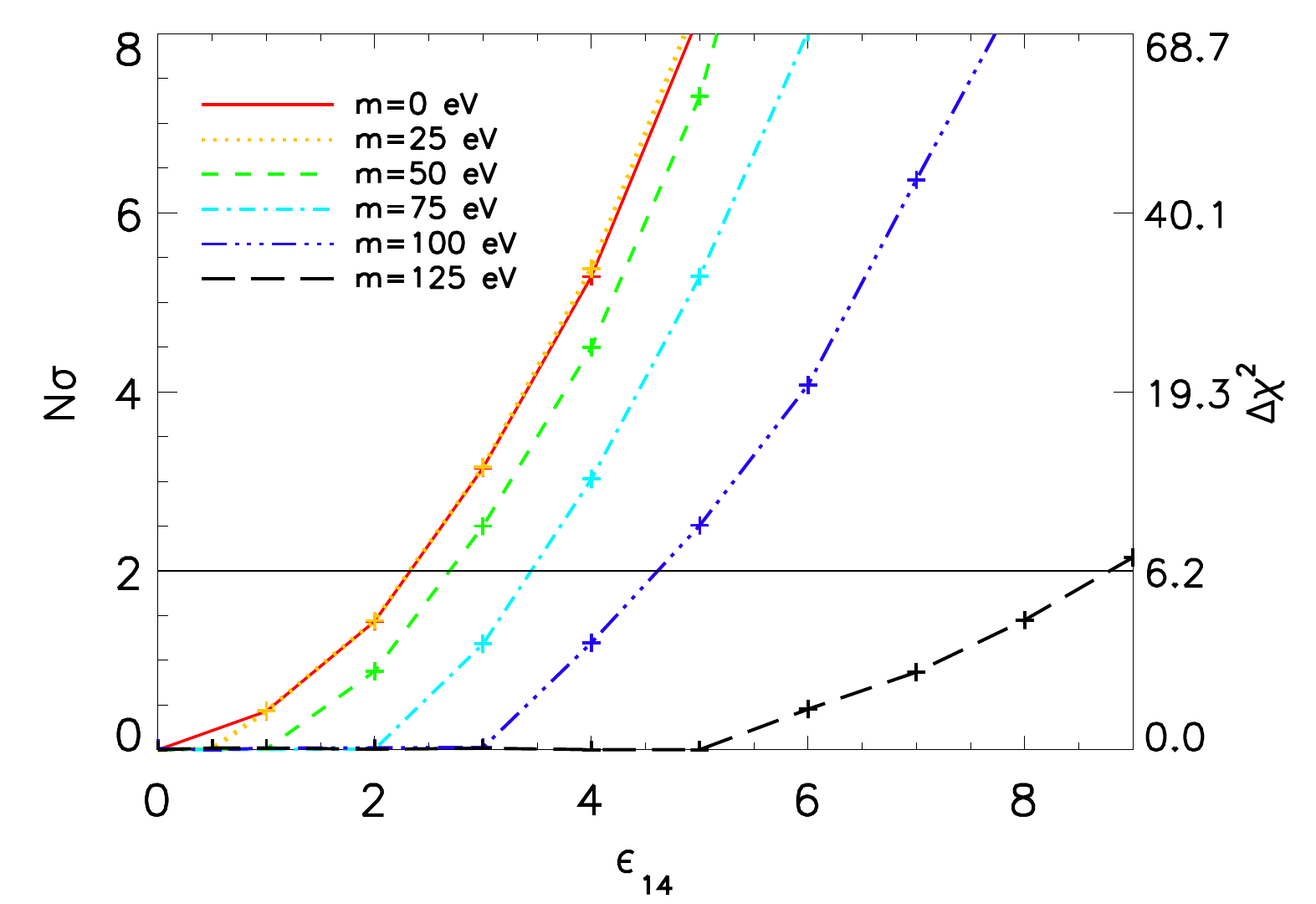}
   	\includegraphics[width=0.49\textwidth]{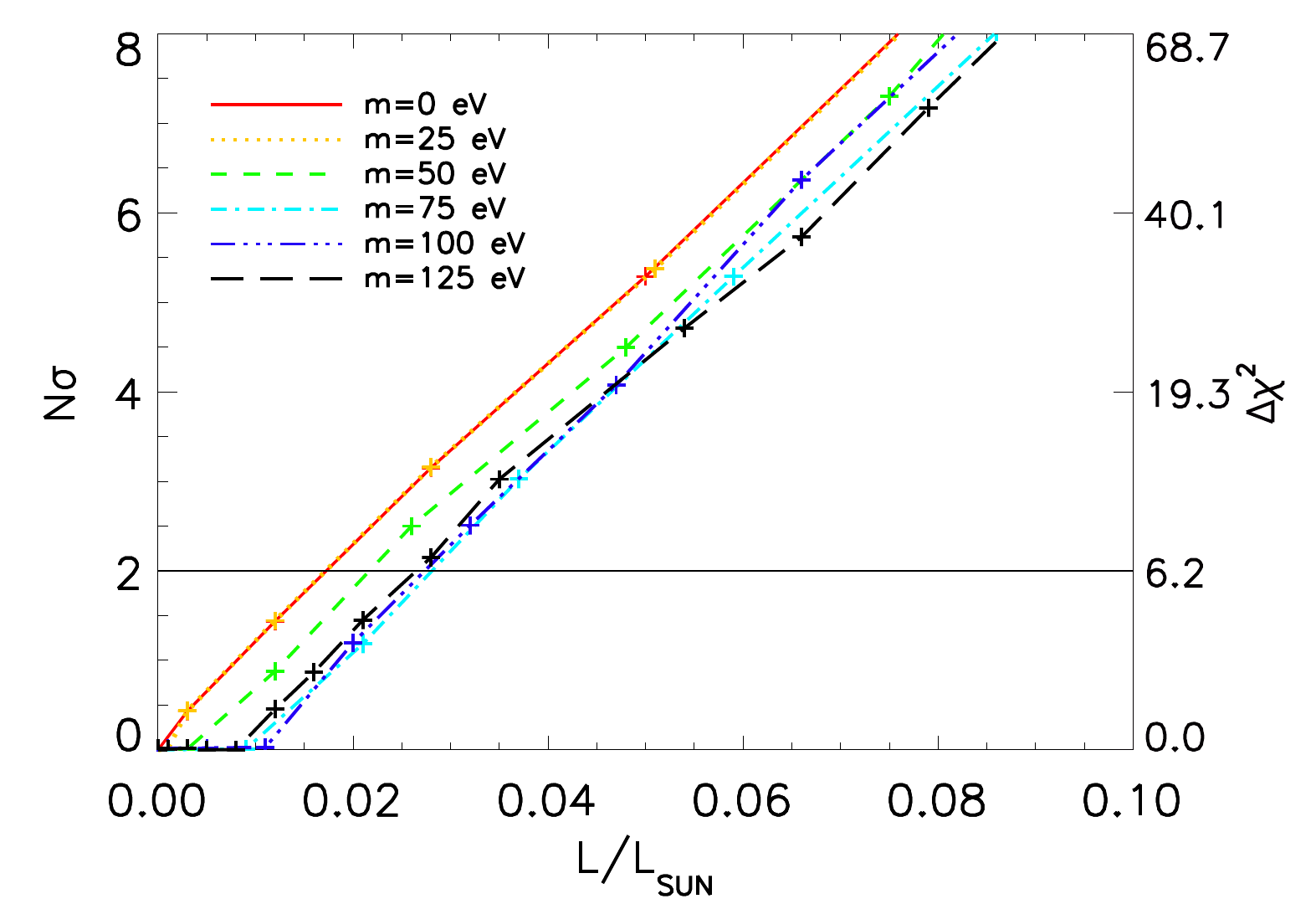}\\
 \caption{ $N \sigma$ and $\Delta \chi^2$ as a function of $\epsilon_{14}$ (left panel) and MCP luminosity $L_{\rm{MCP}}/L_\odot $ (right panel) for models with different $m_f$.}
 \label{fig:chi2}
 \end{figure}

In this paper, we derive an upper limit at $2\sigma$ CL in the ($m_f$, $\epsilon$)-plane. The confidence level can be determined with $\Delta \chi^2 = \chi^2 - \chi^2_{\rm{min}}$, where for a two degrees of freedom problem $\Delta \chi^2 = 2.3, 6.2, 11.8, 19.3$ correspond to  confidence levels of $N\sigma = 1,2,3,4\, \sigma$. In fig.~\ref{fig:chi2}, we plot the resulting values of $N \sigma$ as a function of $\epsilon$ (left panel) and the exotic luminosity (right panel). The bounds at $2 \sigma$ and the corresponding exotic luminosity are summarized in tab.~\ref{tab:results}. 

 \begin{table}[t]
\begin{center}
\begin{tabular}{ |c c c | c c c | } \hline & & & & &\\[-0.35cm]
$m_f (\rm{eV})$ & $\epsilon \times 10^{14}$  at $2\sigma$ & $L_{\rm{MCP}}/L_\odot (\rm\%)$ & $m_f (\rm{eV})$ & $\epsilon \times 10^{14}$  at $2\sigma$& $L_{\rm{MCP}}/L_\odot (\rm\%)$ \\
\hline
0& 2.2& 1.5 &150 & 460 & - \\
25& 2.2& 1.5 & 175& 460 & 2.3 \\ 
50& 2.6& 2.0 & 200 & 500 & 2.4 \\
75& 3.4& 2.7 & 316 & 600 & 2.3\\
100& 4.5& 2.6 & 1000 & 1090 &  2.3\\
125& 8.7 & 2.6 & 3160 & 7720 & 2.8\\\hline
\end{tabular}
\caption{\small{Bounds on $\epsilon$ at $2\sigma$ CL and the corresponding luminosity for different values of $m_f$. For $m_f = 150 \ \rm{eV}$, we take the limit of 
$m_f=175\ \rm{eV}$ (see the main text).}}
\label{tab:results}
\end{center}
\end{table}
 
We find that the  $2 \sigma$ bounds on $\epsilon$ become weaker for larger masses. That is because the MCP emission is reduced and more localized in the inner part as it was shown in fig.~\ref{fig:mcpemission}. However, when we consider this bound as a function of luminosity (right panel of fig.~\ref{fig:chi2}), we can see that for all the cases this luminosity is in a similar range ($L_{\rm MCP}/L_\odot = 1.5 - 2.7 \%$). This dispersion in luminosity at $2\sigma$ is related to the fact that the distribution of energy-loss in the Sun is different for each of the masses which will translate into different effects on the structure of the Sun. That means that we cannot use a general value for the exotic luminosity to constrain the MCPs parameter space because the luminosity depends on the exact emission of the particle-model studied, as it was already mentioned in ref.~\cite{Vinyoles:2015aba}.

While fig.~\ref{fig:chi2} shows the MCP models where $2m_f < \omega_p$ at least in some parts of the Sun so that stringent limits on $\epsilon$ can be achieved, the energy-loss is strongly suppressed everywhere for $2m_f > \omega_p$ where only off-shell emission is present. This suppression will lead to bounds that are much weaker.

Using the statistical method described above, we compute bounds on MCPs until $3160 \ \rm{eV}$. For larger masses, MCPs from the Sun are already disfavored so that it suffices to analytically estimate the behavior of the bound. For $m_f>T$, the emission rate~\eqref{eq:rate2} approximately scales like $m_f\,{\rm exp}(-2m_f/T)$. Conservatively assuming an emission rate of $5\% L_{\odot}$, we obtain the values in tab.~\ref{tab:resultsana} for $m_f>3160$.

Finally, a comment on the model with $m_f=150 \ \rm{eV}$. This case is peculiar because of the two-peaked structure that is present in the center (see fig.~\ref{fig:mcpemission}). The first peak is very narrow which resulted into convergence problems of the solar models. In order to be conservative, for $m_f=150 \ \text{eV}$ we use the bound obtained for the model $m_f=175\ \rm{eV}$ corresponding to a value of $\epsilon = 4.6 \times 10^{-12}$. This is warranted because the emission rate of the $m_f=150 \ \text{eV}$ model is stronger than the emission rate for $m_f=175 \ \text{eV}$ (see fig.~\ref{fig:mcpemission}). The limit on $\epsilon$ would, hence, be stronger for $m_f=150 \ \text{eV}$ than the value adopted here.

\begin{table}
\begin{center}
 \begin{tabular}{ | c c c |}\hline & &\\[-0.35cm] 
$m_f (\rm{eV})$ & $\epsilon$ & $L_{\rm{MCP}}/L_\odot (\rm\%)$ \\
\hline & & \\[-0.35cm]
$10^4$& $1.7\times 10^{-8}$& 5 \\
$10^{4.1}$& $1.3\times 10^{-7}$& 5  \\
$10^{4.2}$& $1.8\times 10^{-6}$& 5 \\
$10^{4.3}$& $4.9\times 10^{-5}$& 5\\
$10^{4.4}$& $3\times 10^{-3}$& 5\\\hline
\end{tabular}
\caption{\small{Analytical estimate for the bound on $\epsilon$ for $m_f> 3160$. We assume an exotic luminosity of $5\%L_{\odot}$.}}
\label{tab:resultsana}
 \end{center}
\end{table}

\section{Discussion and summary}\label{sec:conclusion}

Light, weakly-interacting particles can be well constrained by astrophysical sources. For a global picture, we present in tab.~\ref{tab:summ1} limits on various characteristics of light particles that have been mainly obtained from the energy-loss argument. In the table, we also updated the result for MCPs from the Sun and the constraint on the magnetic moment of the neutrino. In general, constraints derived from the Sun are much weaker than those obtained from red giants, horizontal branch stars, white dwarfs or supernovae because of their more extreme environments. This argument is confirmed by the new limit on the neutrino magnetic moment that we have derived using the same method as for MCPs (see app.~\ref{app:numag}). For the MCPs themselves, however, our result yields a bound similar to those from more massive objects.
\begin{table}[t]\begin{center}\begin{threeparttable}[t]
\begin{tabular}{  |c|c c c|}\hline & & & \\[-0.35cm]
   & MCPs ($\epsilon )$&  L-HP $\kappa m_V [\text{eV}]$ & T-HP $\kappa$\\[0.05cm]\hline & & & \\[-0.35cm]
   Sun &  $2.2\times 10^{-14}$ & $1.8 \times 10^{-12}$ \cite{Vinyoles:2015aba} & $2\times 10^{-15}$~\cite{An:2014twa}\\
   Red Giant & $2\times 10^{-14}$~\cite{Vogel:2013raa,davidson2000}  & $2.1 \times10^{-11}$~\cite{Redondo:2013lna}   & $9\times10^{-16}$~\cite{An:2014twa}\\ 
   Horizontal Branch& $2\times 10^{-14}$~\cite{Vogel:2013raa,davidson2000} & $1\times 10^{-11}$~\cite{An:2013yfc,Redondo:2013lna}
   & $1\times 10^{-15}$~\cite{An:2014twa}\\
   White Dwarf & $1.5\times 10^{-13}$~\cite{davidson2000}  & - & -\\
   Supernova 1987a &  $ 10^{-7}>\epsilon>10^{-9}$~\cite{Mohapatra:1990vq} & -  & -\\
   \hline & & &\\[-0.35cm]
     &  $\mu_\nu/\mu_B$ & $g_{a\gamma}\ [\text{GeV}^{-1}]$ & $g_{ae}$ \\[0.05cm]  \hline & & & \\[-0.35cm]
   Sun &  $1.4 \times 10^{-10}$ (App.~\ref{app:numag}) & $4.1\times 10^{-10}$ ~\cite{Vinyoles:2015aba} & $2.3\times 10^{-11}$~\cite{Redondo:2013wwa} \\
   Red Giant &  $4.5 \times 10^{-12}$ \cite{viaux2014}& $0.66\times 10^{-10}$~\cite{ayala2014} & $4.3\times 10^{-13}$~\cite{viaux2014} \\
   Horizontal Branch&  $1.4\times 10^{-11}$~\cite{raffelt1999} &$0.66\times 10^{-10}$~\cite{ayala2014}& $4.3\times 10^{-13}$~\cite{viaux2014} \\
   White Dwarf  & $5\times 10^{-12}$~\cite{Bertolami:2014noa}
   & $1.5 \times 10^{-11}$~\cite{Gill:2011yp}$^\dagger$ & $2.8\times 10^{-13}$~\cite{Bertolami:2014wua} \\
   Supernova 1987a  & $4\times 10^{-12}$~\cite{Ayala:1998qz}  & $5.3\times 10^{-12}$~\cite{Payez:2014xsa}$^\dagger$ & -\\ \hline
  \end{tabular}\begin{tablenotes} \item \small{$\dagger$ These bounds are not derived directly from the energy-loss argument}\vspace{-0.4cm}\end{tablenotes}
  \end{threeparttable}\end{center}
\caption{Summary of bounds on various light, weakly-interacting particles: The minicharge $\epsilon$ of minicharged particles (MCPs), the product of kinetic mixing and mass $\kappa m_V$ of longitudinal hidden photons (L-HP), and the kinetic mixing $\kappa$ of transversal hidden photons (T-HP), 
the diagonal magnetic moments $\mu_\nu$ of Dirac neutrinos, the ALP-photon coupling $g_{a\gamma}$ and ALP-electron coupling $g_{ae}$ from various astrophysical sources. The bounds for MCPs are valid roughly up to half the plasma frequency of the source. For L-HP the bound becomes less strict already before the mass of the hidden photon reaches the plasma frequency. For T-HP we quote the most stringent values. 
  Limits on neutrino magnetic moments  $\mu_\nu$ were again obtained from plasmon decay. Note that from red giants the best 68\% CL bound has been found in refs.~\cite{Arceo-Diaz:2015pva, Arceo-Diaz:20152}. The presented value for $\mu_\nu$ from the Sun is a new limit obtained by using the same statistical method as for MCPs (see appendix).
   Note that the bound on the ALP photon coupling from white dwarfs does not rely on the cooling argument but on spectral distortion. The bound from the supernova 1987a is derived from the non-observance of a gamma-ray signal in the GRS instrument and is valid for very small masses $m_a<4.4 \times 10^{-10}\ \text{eV}$~\cite{Payez:2014xsa}.}\label{tab:summ1}
\end{table}

To derive this limit, we use the statistical approach described in ref.~\cite{Vinyoles:2015aba} to compare the results of solar simulations with the emission of minicharged particles to models without a contribution from minicharged particles. Using helioseismology and neutrino data, we are able to derive a limit $\epsilon< 2.2\times 10^{-14}\ \text{(95\% CL)}$ on minicharged particles with masses $m_f<25\ \text{eV}$, 
which corresponds to a maximum excess luminosity of $L_{\text{MCP}}/L_{\odot}<1.5\%$. We also computed the behavior of this bound for higher masses of the MCP using on-shell and off-shell plasmon decay. The results are presented in fig.~\ref{fig:MCPresult} giving the most comprehensive summary on the available parameter space for MCPs with and without a hidden photon.

Our low-mass bound ($m_f=0$) is close to the ones derived from red giants and horizontal branch stars. The importance of the Sun for MCPs can be understood by noting that the emission rate per unit mass of MCPs is mostly sensitive to the temperature but not to the density. Moreover, the luminosity constraint for the Sun is more stringent, but also better understood: whereas the usual assumption of $L_{\text{MCP}}/L_{\text{RG, HB}}<10\%$ is rather ad hoc, the result of this paper makes use of the well studied environment of the Sun to yield $L_{\text{MCP}}/L_{\odot}<1.5\%$. This statement also contains statistical information that is not available for other stars.

We would also like to highlight, that for certain parameters of MCP emission, it is possible to increase the neutrino emission of $\rm{^7Be}$ while at the same time decreasing the emission of $\rm{^8B}$. This interesting feature arises, because for large MCP masses the emission is confined to a small region around the center of the Sun.

Since the Sun is our closest astrophysical object, future improvement in our available data and understanding are to be expected. Interestingly, also dark matter experiments are close to being sensitive to tiny fluxes of MCPs from the Sun. The new data to come will allow us to either further constrain the emission of MCPs from the Sun or to ultimately encounter new signs of physics beyond the Standard Model.

\acknowledgments

We are indebted to Aldo Serenelli and Javier Redondo for a series of insightful remarks, and to Georg Raffelt for all the support during this project. NV thanks Francesco Villante and Jordi Isern for their useful advice. HV would like to thank Ignacio Izaguirre for helpful suggestions and Sacha Davidson for a useful correspondence. NV acknowledges funding support from ESP2013-41268-R (MINECO), 2014SGR-1458 (Generalitat de Catalunya) and the MICINN grant AYA2011- 24704. HV acknowledges support by the Deutsche Forschungsgemeinschaft
(DFG) under Grant No. EXC-153 (Excellence Cluster "Universe"),
and by the Research Executive Agency (REA) of
the European Union under Grant No. PITN-GA-2011-289442
(FP7 Initial Training Network “Invisibles”). 

\begin{appendix}
\label{appendix}
\renewcommand{\thetable}{\Alph{table}}
\setcounter{table}{0}
\section{Collider constraints on minicharged particles}
 We updated the collider constraints in fig.~\ref{fig:MCPresult} on minicharged particles including the constraints by both ref.~\cite{Davidson:1991si} and ref.~\cite{Davidson:2000hf}. The values are given in tab.~\ref{tab:coll}.
\begin{table}[t]\begin{center}
 \begin{tabular}{|c c| c c |}\hline
  Mass $m_f$ [GeV] &  minicharge $\epsilon$ &   Mass $m_f$ [GeV] &  minicharge $\epsilon$  \\\hline
  $<84$  & $>2/3$ & $<0.1$  & $>0.01$\\
  $<45$  & $>0.24$ & $<0.01$  & $>3\times 10^{-3}$\\
  $<13$  & $>0.2$ & $<2\times10^{-3}$  & $>10^{-3}$\\
  $<10$  & $>0.09$ & $<1\times10^{-3}$  & $>6\times10^{-4}$\\
  $<5$  & $>0.08$ & $<2\times10^{-4}$  & $>3\times10^{-4}$\\
  $<1$  & $>0.03$ & & \\\hline
 \end{tabular}
 \end{center}\label{tab:coll}
 \caption{Collider constraints according to refs.~\cite{Davidson:1991si,davidson2000}.}
\end{table}
 
  \section{Neutrino magnetic moments}\label{app:numag}

 Using the same method as for MCPs, we can also derive a bound on neutrino magnetic moments. The emission rate of massless neutrinos via plasmon decay is again given by
\begin{align}\label{eq:burate}
 \mathcal{Q} = \frac{2}{2\pi^2}\int_0^\infty \dd k k^2\frac{ \omega \Gamma_{\nu}}{e^{\omega/T}-1}\,,
\end{align}
where the decay rate is~\cite{Raffelt:1996wa}
\begin{align}\label{eq:nu2}
 \Gamma_{\nu}=\frac{\mu_\nu^2}{24\pi} \frac{Z}{\omega}\omega_p^4\,.
\end{align}
Here $\mu_\nu$ is the neutrino magnetic moment, which we want to constrain. Note that the rate depends on $\omega_p^4$ instead of $\omega_p^2$ as was the case for MCPs. This stronger dependence of the density makes the Sun a weaker probe than the more dense RGs and HB stars. 

The obtained bound is
\begin{equation}
 \mu<1.4 \times 10^{-10}\, \mu_B\,,
\end{equation}
which corresponds to a luminosity of $L_{\mu_\nu}<1.3\% \, L_\odot$. This limit is better than previous estimates from the Sun~\cite{raffelt1999} but worse than the values obtained from RGs in ref.~\cite{viaux2014}.
 
\end{appendix}

\end{document}